# Cost-based variable selection for prediction models

Lasai Barreñada[1,2], Andrew Vickers[3], Laure Wynants[1,2,4], Bieke Tack[5], Dirk Timmerman[1,6], Ewout Steyerberg[7], Ben Van Calster[1,2,7]

[1] Department of Development and Regeneration, KU Leuven, Belgium

[2] Leuven Unit for Health Technology Assessment Research (LUHTAR), KU Leuven, Belgium

[3] Department of Epidemiology and Biostatistics, Memorial Sloan Kettering Cancer Center, New York, NY, USA

[4] Department of Epidemiology, CAPHRI Care and Public Health Research Institute, Maastricht University, Maastricht, Netherlands

[5] Department General Internal Medicine, University Hospitals Leuven, Leuven, Belgium

[6] Department Obstetrics & Gynecology, University Hospitals Leuven, Leuven, Belgium.

[7]Julius Centre for Health Sciences and Primary Care, University Medical Centre Utrecht, Utrecht, Netherlands

**Corresponding author**

Ben Van Calster

KU Leuven, Department of Development and Regeneration

Herestraat 49 box 805 T +32 16 377 788

3000 Leuven

Belgium

ben.vancalster@kuleuven.be

**Keywords: Variable selection, prediction model, net benefit, feature selection, clinical utility**

**Objective:** Variable selection for clinical prediction models is commonly done on statistical grounds related to model fit. Statistical approaches ignore the cost of misclassifications and the harm of measuring a variable. We aim to develop a decision-analytic variable selection algorithm based on net benefit.

**Materials and Methods:** The proposed algorithm selects the set of predictors by exhaustively comparing the net benefit of all combinations using cross validation, with an extension based on backward elimination for settings with many predictors. Variables are selected if the increased net benefit from better predictiveness exceeds the net benefit reduction from the harm of variable measurement. Harm includes financial expense, time, resource allocation, and patient burden. We illustrate the algorithm through a simulated example and three real case studies. We compare its performance to traditional backward elimination and LASSO.

**Results:** Across the case studies, parsimonious models were generated that optimized clinical utility with minimal impact on discrimination and calibration. Penalizing predictors based on their cost ensured that expensive or burdensome variables were only selected if they offer sufficient clinical utility. The algorithm was implemented in the R package *NBvarsel*.

**Discussion:** Decision-analytic variable selection allows prediction models to maximize patient benefit while minimizing unnecessary clinical burden and costs.

**Conclusion:** Clinical utility should not be an afterthought reserved only for implementation. By integrating a harm-adjusted net benefit metric directly into the variable selection process, *NBvarsel* allows researchers to select only the variables that sufficiently improve clinical utility.

# 1 BACKGROUND AND SIGNIFICANCE

Clinical prediction models are statistical tools that estimate the probability that a patient has or will develop a condition in the future [1]. These models try to link the outcome (event or condition) with a combination of predictors in a dataset [2]. These predictors should be readily available at the time of prediction, not too costly to measure, and measured with reasonable precision [1]. However, there might be a large number of predictors that meet these criteria such that selection of a subset of predictors is required [3].

Variable selection, feature selection, and feature engineering are terms used in prediction modelling to describe the process of selecting which predictors to use when training a model. Although a-priori or domain knowledge is highly recommended, data-driven approaches are often needed . Different approaches for variable selection have been presented [4–6]. A distinction can be made between filter, wrapper and embedded methods. Filter methods assess each variable separately by quantifying the association with the outcome. Wrapper methods fit many models using different variable subsets to find the optimal subset. Examples include forward selection, backward elimination, stepwise selection, best subset selection, and recursive feature elimination. Embedded methods select variables while fitting the model. Examples include lasso, ridge, elastic net, or tree-based algorithms like XGBoost. This work focuses on wrapper methods.

To be used in clinical practice, prediction models need to show good performance in external validation in terms of discrimination (area under the ROC curve), calibration (calibration plot), and clinical utility (Net Benefit, NB) [7]. Utility values, relating to misclassification costs (i.e., the benefit of treating the diseased and the harm of treating the healthy), are hard to elicit at an absolute scale. However, it is known that the ratio of harm to benefit is directly related to the decision threshold for classifying a patient as high risk [8]. NB uses this relationship and is calculated as benefit minus harm scaled by the harm-to-benefit ratio [9]. In statistical terms this is the proportion of true positives minus the proportion of false positives weighted by the harm-to-benefit ratio, which reduces to the net proportion of true positives. The harm-to-benefit ratio reflects how many false positives a clinician is willing to accept per true positive. For example, suppose that a surgeon is willing to perform at most 10 biopsies to catch 1 malignant tumor. This corresponds to performing a biopsy when the risk of malignancy is ≥10% (the decision threshold). Because the surgeon accepts up to 9 false positives (biopsies with benign findings) per true positive, the harm-to-benefit ratio is 1/9. Given the variability in the preferred harm-to-benefit ratio across clinicians and settings, NB is often shown for a range of acceptable decision thresholds [10]. If clinical utility is important for prediction models, we argue that clinical utility should already be addressed at the variable selection stage.

To illustrate the issue using a simple motivating example, imagine two prediction models for cancer, where a patient at high risk is referred for biopsy. The “clinical” model only uses routinely available clinical variables, the “imaging” model also includes the results of an expensive, inconvenient and unpleasant imaging test. Even if the imaging model had better discrimination, the improvement may not outweigh the added costs and harms of the imaging test. This can be evaluated by including test harm into the NB formula. Then, the clinical model may still have higher utility despite having worse discrimination.

Now imagine no models were available to predict the result of biopsy and a research group therefore decided to build one. Using the conventional approach of maximizing discrimination, they might well develop a model that includes the imaging test, an approach that would be inferior on clinical implementation. We propose clinical utility can often be usefully incorporated into model building rather than being used only at the evaluation stage.

# 2 OBJECTIVE

The aim of this paper is therefore to develop a simple and interpretable variable selection algorithm based on clinical utility to determine which predictors to include in the model and implement it as an easy-to-use R package. We study the impact on performance through a simulation and three real case studies. We will use and implement the method in the context of a logistic regression to estimate the probability of a binary outcome, but the rationale extends to more flexible machine learning or AI probabilistic algorithms.

# 3 MATERIALS AND METHODS

## 3.1 NET BENEFIT AND TEST HARM

Net benefit is defined as: $NB = \frac{TP}{N} - \frac{FP}{N} \times \frac{p_t}{1-p_t}$ where TP and FP are true positives and false positives, respectively and $p_t$ is the decision threshold associated with the harm-to-benefit ratio. Test harm is related to the cost that measuring a predictor (i.e. conducting a test) has. This can be defined as the reciprocal of the number of tests that a doctor would be willing to do assuming that the test is perfect [11]. We define a harm adjusted NB as NB minus the test harm**.** Harm is therefore accounted for twice. First, to calculate NB, we set a threshold that should be related to the **harm or cost of misclassifying a patient**. Second, the **test harm** is introduced in the NB scale by quantifying the minimal expected benefit required to measure a variable (**Box 1**). The harm of misclassifying a patient is widely discussed in the literature [9,10]. Imagine we have a model that predicts the risk that an ovarian tumor is malignant based on age and ultrasound measurements of the lesion. At a 10% decision threshold the NB is 0.2. We wonder whether measuring CA125 improves clinical utility. CA125 requires a blood sample, a processing cost, and a delay until the results are known. Given these costs, practitioners indicate that they would be willing to measure CA125 in at most 80 patients to find an event. The test harm is then $\frac{1}{80} = 0.0125$. We would be willing to include the predictor only if the NB increases by at least 0.0125. Other predictors may incur lower or higher costs. A predictor based on CT-scan results, for instance, has a higher financial cost and also causes discomfort and exposure to radiation. In contrast, predictors such as age or sex have no costs and therefore a test harm of 0 such that any improvement in NB justifies their inclusion.

| **Box 1: Test harm** |
|---|
| We acknowledge that determining the harm of measuring a predictor in the context of a model development study is challenging. In this box we propose some aspects to consider when determining the harms.<br>- **Harms are broad:** They go beyond monetary costs of a test, including patient burden, possible side effects, objectivity of the measurement, requirement of specific equipment for the measurement, time needed to undergo the test, etc.<br>- **Harms are population dependent:** A predictor might be very costly for the general population, but in the context of the prediction model problem be readily available, because it is measured as part of the clinical routine in the target population. |

- **Combined or conditional harms:** Some procedures, such as ultrasound examinations, provide data on multiple predictors at once. Consequently, the assigned harm of measuring a single predictor from this procedure can be considered equal to the harm of measuring all predictors based on the procedure. This illustrates the concept of conditional harm. For instance, once an ultrasound is underway, the additional harm of recording a second predictor is zero. However, conditional harms are not always zero. Measuring a primary biomarker requires both a blood draw and laboratory processing. Measuring a second biomarker from the same sample conditionally avoids the physical harm of another blood draw, but still incurs the harm of additional lab processing.

In the algorithm presented the harms should be defined according to these aspects.

## 3.2 ALGORITHM

The problem we aim to address is defined as follows: given a set of predictors $P = \{p_1, p_2, \ldots, p_P\}$ and a modelling algorithm $A$ we seek the subset $S \subseteq P$ that maximises the net benefit:

$$S = \arg\max_{S \subseteq P} NB(S, A, p_t)$$

Here, $NB(S, A, p_t)$ denotes the (adjusted) net benefit of algorithm $A$ with $S$ predictors for the decision threshold $p_t$. The $p_t$ parameter may be a single value, or a range of clinically relevant values specified based on clinical context. In cases where a range of values is used, we maximize the average NB across that range.

The proposed variable selection algorithms (**Algorithms 1-2**) do not involve a modification of the coefficient estimation procedure, since the logloss or cross-entropy minimisation is known to provide coefficients with good properties [12].

**Algorithm 1: The exhaustive NB variable selection (NBvarsel)**

1. Define all possible $K$ model combinations based on the $P$ predictors: $K = 2^P - 1$.
2. Assign test harms to adjust the NB accordingly. If the harm is unknown, this can be 0 and unadjusted NB will be used.
3. Train $K$ models and evaluate their net benefit using cross validation with $f$ folds (default $f$ is 5).
    a. This step can be parallelized to reduce computation time.
    b. Nonlinearities can be addressed with restricted cubic splines.
    c. Interactions can be included; we limit this option to either all pairwise interactions or no interactions.
4. Rank the $K$ models based on NB and select the model with optimal NB.

**Algorithm 1** becomes computationally too demanding when the number of candidate predictors increases. For a dataset with 20 candidate predictors, there are more than one million predictor combinations; consequently, the method fits 5 million models when using 5-fold cross-validation. In most trials or observational studies, the sample size is limited; therefore, the number of candidate predictors should also be kept small for which a priori variable selection is recommended. However, in those cases where there are many candidate predictors (e.g. electronic health records), we propose an extension of the methodology that we called backward elimination NB variable selection (**Algorithm 2**).

| Algorithm 2: The backward elimination NB variable selection (BE NBvarsel) |
|---|
| 1. Assign test harms.<br>2. Train the model with all predictors ($P$).<br>3. Define all possible $K$ model combinations based on $P$ and number of predictors per step $s$ by removing s predictors from the complete model: $K = \binom{P}{s}$.<br>4. Train the models from steps 2 and 3 and evaluate their net benefit using cross validation with $f$ folds. This step includes the same functionalities as **Algorithm 1**.<br>5. Rank the models based on NB.<br>a. If the model with best NB is the model obtained in step 2 the variables included in that model are selected and the algorithm stops.<br>b. If not, the predictors from the model that presented the lowest NB are removed from $P$ and steps 2-5 are repeated. |

Algorithms 1 and 2 should be used according to modern sample size recommendations for prediction models, which calculate the minimum required participants based on anticipated model performance and complexity, including as parameters all candidate predictors [13]. The algorithm allows calculating variable importance using permutation (details **in Supplementary Material S1**). Code implementing the NB based variable selection and the associated visualizations are available in the open-source R package *NBvarsel* **available on CRAN (https://lasaibarrenada.github.io/NB_varsel/)** [14]**.**

## 3.3 MOTIVATING EXAMPLE

To illustrate the methodology consider the biopsy risk prediction model mentioned in the introduction. This model incorporates a cost-free clinical variable, a blood biomarker (test harm = 1/100), and a costly, burdensome imaging test (test harm = 1/20). The prevalence of cancer is 22%. Standard statistical selection retains all predictors, yielding an AUC of 0.75. However, cost-adjusted decision curves (**Figure 1A**) reveal that this full model provides no greater utility than a "treat all" strategy, when test harms are accounted for. Conversely, our proposed NB-guided selection, using thresholds from 15% to 40%, excludes the costly imaging predictor, although discrimination decreases (AUC = 0.67; **Table 1**). The simpler model achieves clinical utility at risk thresholds above 15%. The all-subset plot included in the NBvarsel package (**Figure 1B**) effectively visualizes this tradeoff, showing that adding the imaging test decreases overall NB because its high cost outweighs its marginal predictive gain. Methodological details of the simulation for the motivating example are available in **Supplementary Material S2**.

**Table 1. Results for the motivating example. $X_{clin}$:clinical predictor, $X_{img}$: imaging predictor, $X_{blood}$: blood biomarker. Sorted by average adjusted net benefit.**

| Included predictors | Total Cost | Avg. Net Benefit | Avg. Adj. Net Benefit | AUC | Brier Score |
|---|---|---|---|---|---|
| $X_{blood}, X_{clin}$ | 0.01 | 0.040297 | 0.030297 | 0.669017 | 0.162685 |
| $X_{clin}$ | 0 | 0.023164 | 0.023164 | 0.600359 | 0.170146 |
| $X_{blood}$ | 0.01 | 0.031515 | 0.021515 | 0.638132 | 0.166590 |
| $X_{img}, X_{blood}, X_{clin}$ | 0.06 | 0.067989 | 0.007989 | 0.754671 | 0.147707 |
| $X_{img}, X_{clin}$ | 0.05 | 0.054519 | 0.004519 | 0.714418 | 0.155900 |
| $X_{img}, X_{blood}$ | 0.06 | 0.061838 | 0.001838 | 0.735281 | 0.151690 |
| $X_{img}$ | 0.05 | 0.046665 | -0.003335 | 0.690021 | 0.159453 |

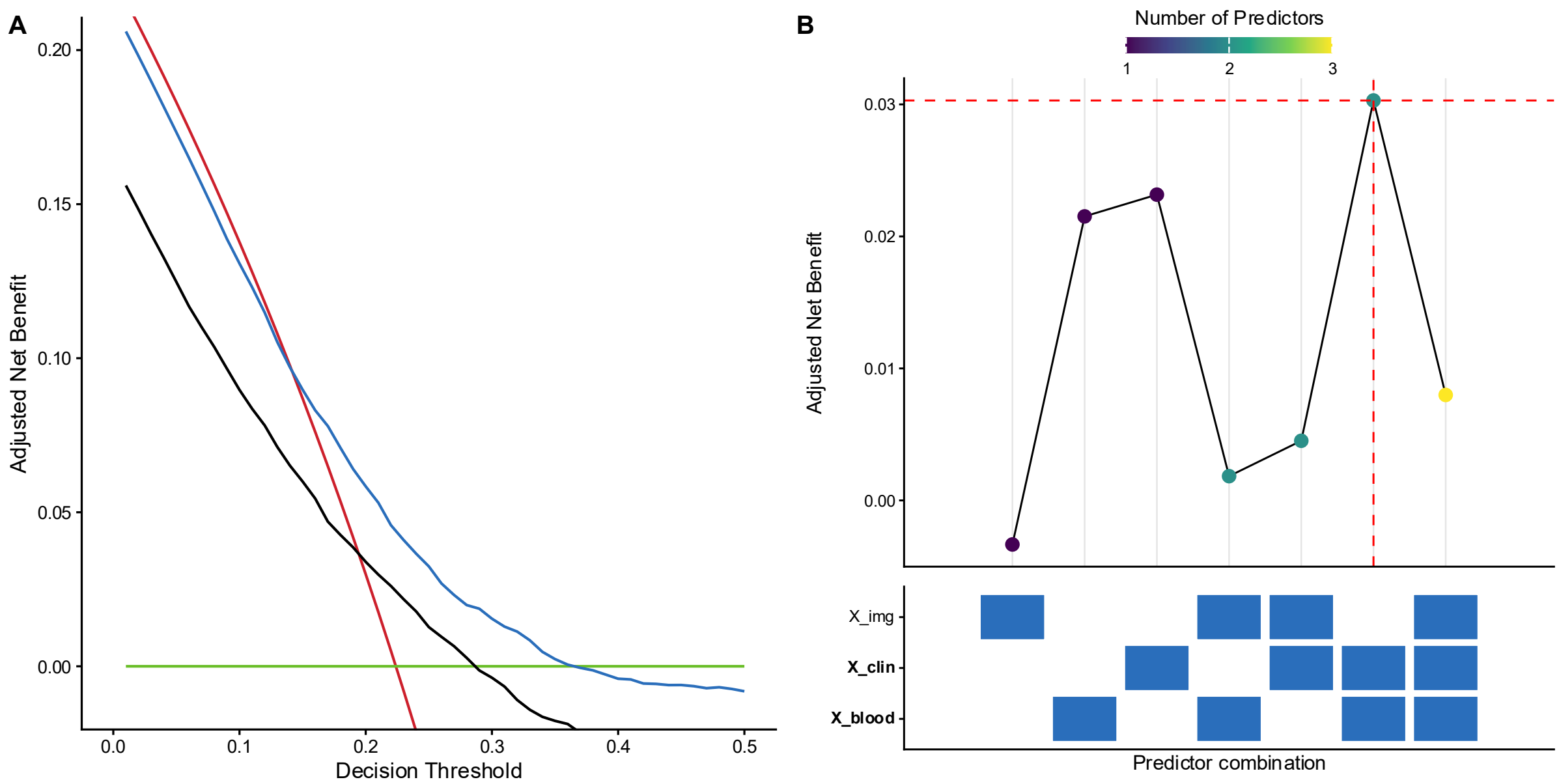

**Figure 1. (A) Cost adjusted decision curve analysis for the motivating example for NBvarsel model (blue), full model (black), treat all (red) and treat none (green) . (B) All subset plot for the simple illustration. Predictors in bold optimize adjusted net benefit.**

# 4 CASE STUDIES

We will test the variable selection methodology (NBvarsel) with three datasets previously used to develop clinical prediction models. We will then compare the performance in terms of discrimination (AUC), calibration (calibration plots and the ratio of observed to expected events; OE ratio) and clinical utility (NB and DCA) of NBvarsel, backward elimination based on AIC (equivalent to p-value selection with $\alpha = 0.157$) and LASSO regularization. For LASSO, the penalty parameter ($\lambda$) is tuned via cross-validation to select the model that yields the minimum cross-validated error. All case studies fulfill the minimum sample size requirements for model development calculated using Riley's criteria [13] (**Supplementary report).**

All the analyses were performed using R version 4.4 with NBvarsel package version 0.2. We share complete R code to replicate all and additional results presented in this paper in a **Supplementary Report**
(https://github.com/LasaiBarrenada/NB_varsel/blob/v0.1.2/NBvarsel_supplementary_report.html).

## 4.1 OVARIAN CANCER DIAGNOSIS – IOTA DATA

The optimal management of patients with an ovarian tumour depends on the histology of the mass. Benign masses do not always need to be removed and can be monitored with ultrasound follow up [15]. In 2014, the ADNEX model was published using data from 5909 patients (1929 events) [16]. The model is widely used and has shown excellent performance across different settings and populations [17,18]. ADNEX is a multinomial model that predicts the risk of benign, borderline, stage I invasive cancer, stage II-IV invasive cancer or metastatic cancer. In this work, we will treat it as a binary model, where the risk of malignancy is equal to the sum of the four malignant subtypes. As an external test dataset we will use data from the IOTA phase 5 study (N = 6847, events = 1713) [19,20].

To develop ADNEX an a priori selection of 10 predictors based on domain knowledge was used followed by backward elimination selecting 9 predictors of which three were clinical and six were ultrasound-based predictors [16]. The removed predictor was family history of ovarian cancer. For our analysis, missing CA125 values were handled via a regression-based imputation on the training data using all other ADNEX predictors, following recent methodological recommendations [21]. Then this model was applied to the training and test dataset to impute missing values. In the test dataset, some of the tumours were classified as uncertain for various reasons [19]. In the published study, these values were imputed with multiple imputation. We use one set of those imputed values for our analysis.

We extended the original set of 10 predictors with six additional features. The distribution of predictors by outcome for the training and test dataset is included in **Table S1**. In total 65,536 predictor combinations were evaluated across decision thresholds from 1% to 20% risk with 20-fold cross validation. The recommended risk for presurgical diagnosis of ovarian tumours is 10% but we extended it to accommodate different preferences [22]. We optimized the average adjusted NB across those thresholds. Adjusted NB was calculated with harms based on three groups of variables: clinical history variables (test harm = 0, should be available for all patients from the records), ultrasound-based variables (test harm = 1/200), blood-based biomarker (test harm = 1/80).

### 4.1.1 Results

All models except LASSO effectively removed family history of ovarian cancer. NBvarsel models removed maximum diameter of the solid part of the tumour and in addition the adjusted NBvarsel removed CA125 biomarker (**Table 2)**. According to the variable importance calculations based on NB, the least important predictor was family history of ovarian cancer, and the most important one was the proportion of solid tissue (**Figure S1**).

The models showed comparable performance in general, with BE and NBvarsel approaches outperforming slightly the rest **(Table 2)**. All models tended to underestimate the risk of malignancy as shown in the clustered calibration plots with the backward elimination method yielding a slightly closer curve to the diagonal (**Figure S2**). Clinical utility of all models was comparable except for ADNEX with CA125 and LASSO that showed slightly lower utility, as shown in the decision curves **(Figure 2, left)**, but when the costs of measuring a predictor are accounted for, the adjusted NBvarsel and ADNEX without CA125 had higher NB across the whole range of decision thresholds due to removing the costly blood marker **(Figure 2, right)**. All-subset plots are available in Supplementary material (**Figures S3-S4**)

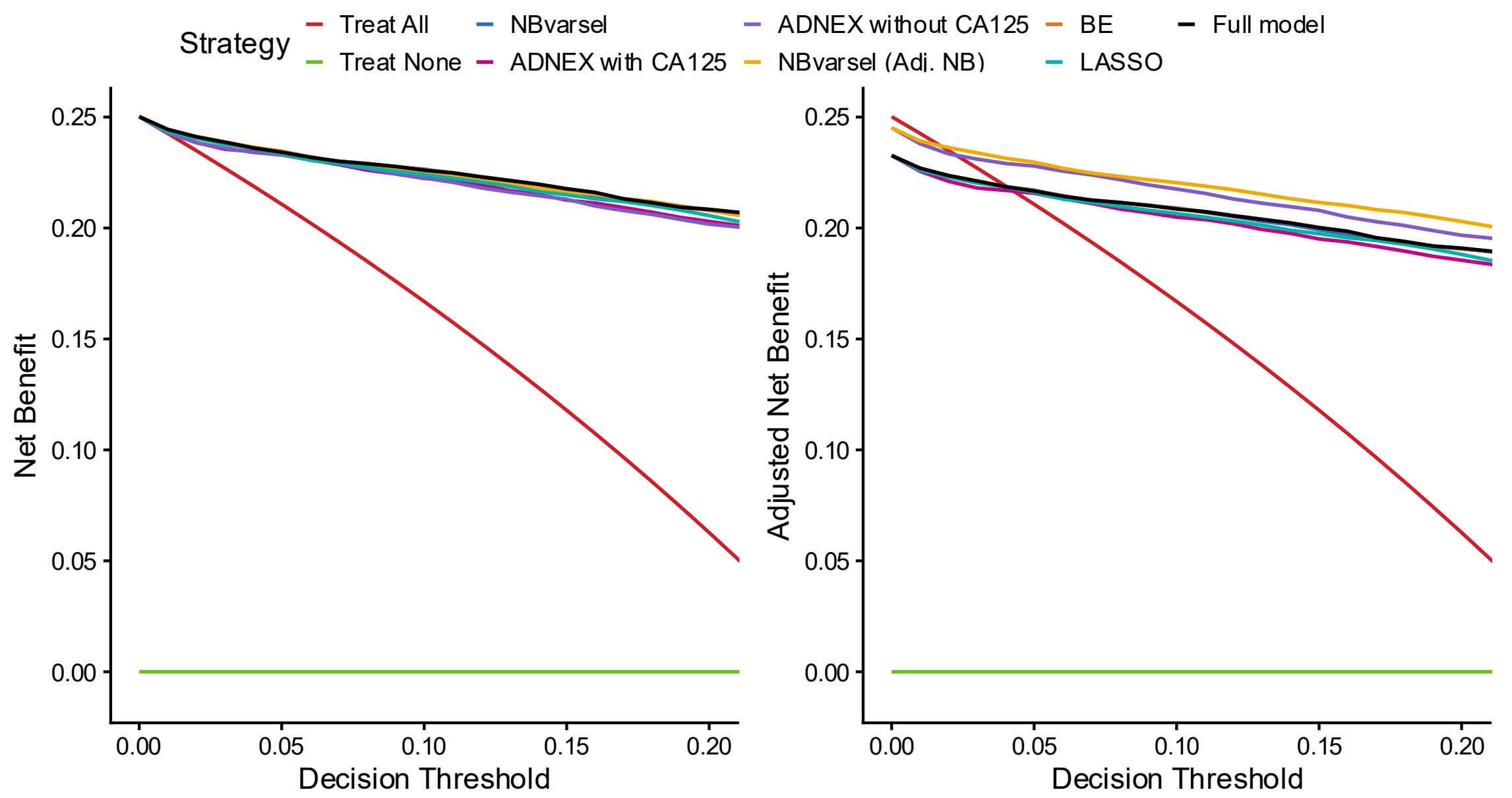

**Figure 2. Decision curve analysis for unadjusted (left) and harm-adjusted (right) for the different variable selection procedures.**

**Table 2. Performance metrics of the different models in the test set varying the variable selection procedure.**

| **Model** | **Removed predictors** | **# predictors removed** | **AUC** | **O:E ratio** | **Brier** | **Avg. NB[a]** | **Cost** | **Avg. Adj. NB[a]** |
|---|---|---|---|---|---|---|---|---|
| NBvarsel | Family history of ovarian cancer<br>Maximum diameter of the solid part of the lesion | 2 | 0.961 | 0.975 | 0.063 | 0.225 | 0.018 | 0.207 |
| NBvarsel (cost adj.) | Family history of ovarian cancer<br>Maximum diameter of the solid part of the lesion<br>CA125 biomarker | 3 | 0.959 | 0.975 | 0.065 | 0.225 | 0.005 | 0.220 |
| LASSO | None | 0 | 0.955 | 0.940 | 0.069 | 0.223 | 0.018 | 0.206 |
| BE | Family history of ovarian cancer | 1 | 0.961 | 0.957 | 0.064 | 0.225 | 0.018 | 0.208 |
| Full model | None | 0 | 0.961 | 0.956 | 0.064 | 0.225 | 0.018 | 0.208 |

NB: Net Benefit. BE: Backward elimination. O:E: Observed over expected.

[a]NB values are averaged over thresholds 1 to 20%.

## 4.2 EMERGENCY DEPARTMENT HOSPITALIZATION – MIMIC-IV ED

For the second case study we will use the MIMIC-IV emergency department (MIMIC-IV-ED) dataset combined with the MIMIC-IV to predict the risk of hospitalization after visiting the emergency department [23–25]. MIMIC-IV-ED contains electronic health records data collected retrospectively from more than 425,000 emergency department stays in the Beth Israel Deaconess Medical Center in Boston, between 2011 and 2019. We use the benchmark proposed by Xie et al., but with the latest version of MIMIC-IV (v3.1) and MIMIC-ED (v2.2) [26]. The authors propose this benchmark to allow different model developers to make fair comparisons when using MIMIC-IV data, therefore we followed the same procedure to obtain the training and test data according to their guidelines (see https://github.com/nliulab/mimic4ed-benchmark/tree/main). The training data contains 334,420 (80%) ED visits and the testing data 83,605 (20%).

In the benchmark, the authors use 64 predictors to develop models to predict risk of hospitalization. No test harms were defined. They use different algorithms, but we will focus on the logistic regression model, which we denote as full model. Hospitalization is defined as the patient being admitted to an inpatient care site immediately after the visit to the emergency department. The incidence of hospitalization was 47%. The predictor distribution across test and training datasets is identical and displayed in **Table S2**. If we were to use the exhaustive approach as in the previous case this would mean approximately $1.84 \times 10^{19}$(million trillion) combinations, which is not feasible. Therefore we will use the BE NBvarsel approach designed for settings with many variables. We set group size to one and two. We used the algorithm with 5 cross validations. We evaluated decision thresholds between 50% and 70% linked to a scenario where the model is used for early resource allocation, such as reserving an inpatient bed. Given acute hospital resource constraints, a false positive (allocating a bed to a patient who ultimately does not need one) can be considered operationally more harmful than a false negative (failing to reserve a bed early for a patient who does).

### 4.2.1 Results

For the MIMIC IV-ED dataset, 15 predictors were removed using this algorithm. The method with group size of two iteratively removes predictors in pairs. This algorithm removed 16 predictors in 8 steps. Backward elimination removed 16 predictors, and LASSO 13.

The most important predictors according to BE NBvarsel were the acuity at triage (Emergency Severity Index), age of the patient and ED and hospitalization in the previous year (**Figure S5**). The most important variables for BE NBvarsel were similar between group size of one and two and also to those reported in the original study calculated using random forest variable importance [26]. All approaches obtained similar AUCs (0.812). Calibration was identical in all models (**Figure S6**). BE NBvarsel models and BE typically ranked slightly higher for clinical utility across the decision curve except for very high thresholds where the full model showed higher utility (**Table 3**).

**Table 3. Net benefit performance in test data for MIMIC-ED comparison. Green cell indicates the best approach in terms of Net Benefit.**

| Decision Threshold | Treat All | Treat None | BE NBvarsel 1 P = 49 | BE NBvarsel 2 P = 48 | Full model P = 64 | BE P = 48 | LASSO P = 51 |
|---|---|---|---|---|---|---|---|
| 0.05 | 0.446558 | 0 | 0.446994 | 0.446979 | 0.446701 | 0.446959 | 0.446993 |
| 0.10 | 0.415811 | 0 | 0.419875 | 0.419827 | 0.419434 | 0.419859 | 0.419911 |
| 0.15 | 0.381447 | 0 | 0.393107 | 0.393126 | 0.391698 | 0.393181 | 0.392990 |
| 0.20 | 0.342788 | 0 | 0.368196 | 0.368220 | 0.366575 | 0.368273 | 0.368199 |
| 0.25 | 0.298973 | 0 | 0.343548 | 0.343496 | 0.342252 | 0.343464 | 0.343608 |
| 0.30 | 0.248900 | 0 | 0.318094 | 0.318060 | 0.318014 | 0.318135 | 0.318060 |
| 0.35 | 0.191123 | 0 | 0.291783 | 0.291626 | 0.291323 | 0.291616 | 0.291644 |
| 0.40 | 0.123717 | 0 | 0.264123 | 0.263908 | 0.263816 | 0.264183 | 0.263959 |
| 0.45 | 0.044055 | 0 | 0.236210 | 0.236240 | 0.235611 | 0.236352 | 0.236180 |
| 0.50 | -0.051540 | 0 | 0.210083 | 0.209904 | 0.208504 | 0.209964 | 0.209760 |
| 0.55 | -0.168378 | 0 | 0.181922 | 0.181573 | 0.179455 | 0.182166 | 0.181831 |
| 0.60 | -0.314425 | 0 | 0.154375 | 0.154303 | 0.151390 | 0.154441 | 0.153747 |
| 0.65 | -0.502200 | 0 | 0.126489 | 0.126570 | 0.124094 | 0.126359 | 0.126561 |
| 0.70 | -0.752567 | 0 | 0.096266 | 0.096174 | 0.095429 | 0.096330 | 0.096286 |
| 0.75 | -1.103080 | 0 | 0.069075 | 0.069027 | 0.067508 | 0.068357 | 0.068226 |
| 0.80 | -1.628850 | 0 | 0.041445 | 0.041277 | 0.040344 | 0.041553 | 0.041576 |
| 0.85 | -2.505133 | 0 | 0.016626 | 0.017204 | 0.019321 | 0.017511 | 0.017383 |
| 0.90 | -4.257700 | 0 | 0.002787 | 0.001531 | 0.006495 | 0.001208 | 0.002476 |
| 0.95 | -9.515400 | 0 | -0.006339 | -0.004904 | 0.000167 | -0.006327 | -0.005418 |

## 4.3 Non typhoidal Salmonella (NTS) - DeNTS study

The last case study used a prospective cohort to predict the risk of non-typhoidal Salmonella (NTS) bloodstream infection in children under five to guide the choice of empirical antibiotics [27]. The dataset contains clinical information collected from 2,327 paediatric hospital admissions with severe febrile illness at the Kisantu district hospital in the Democratic Republic of Congo (**Table S3**). 2323 cases with complete data were used in the analysis as proposed in the original article. We used the a priori clinical predictors proposed by Tack et al., applying their exact variable definitions and restricted cubic spline transformations for continuous predictors. The authors propose this specific set of predictors based on a systematic literature review to ensure clinical relevance and avoid data-driven bias. All predictors included are routinely available in the setting; therefore, no test harm was defined.

In the published work, variables were selected using a double backward elimination process and the final logistic regression model was fitted using Firth's correction [28]. We will compare their approach to the introduced NBvarsel including splines for continuous predictors and then fitting the final model also using Firth's correction. In our algorithm, continuous predictors were introduced either with splines or were removed. The authors propose as clinically relevant range of decision thresholds to be between 20% and 50%, so these were the range of thresholds used in the variable selection procedure. No test data were available; therefore, we present optimism corrected results using 1000 bootstraps comparing the different strategies and our algorithm. Apparent results are included as **Supplementary material S3**.

### 4.3.1 Results

Age at admission and nutritional status were selected in over 50% of the bootstrap runs by the different strategies yet were picked up by NBvarsel only 10% and 33% of the time, respectively. For the remaining predictors, selection frequencies consistently fell either strictly above or strictly below 50% across the various strategies except LASSO (**Figure S12**).

Discrimination and calibration metrics were similar for all models (**Table S4**). However, the median NB across the reasonable range of decision thresholds was slightly superior for NBvarsel (**Figure 3**).

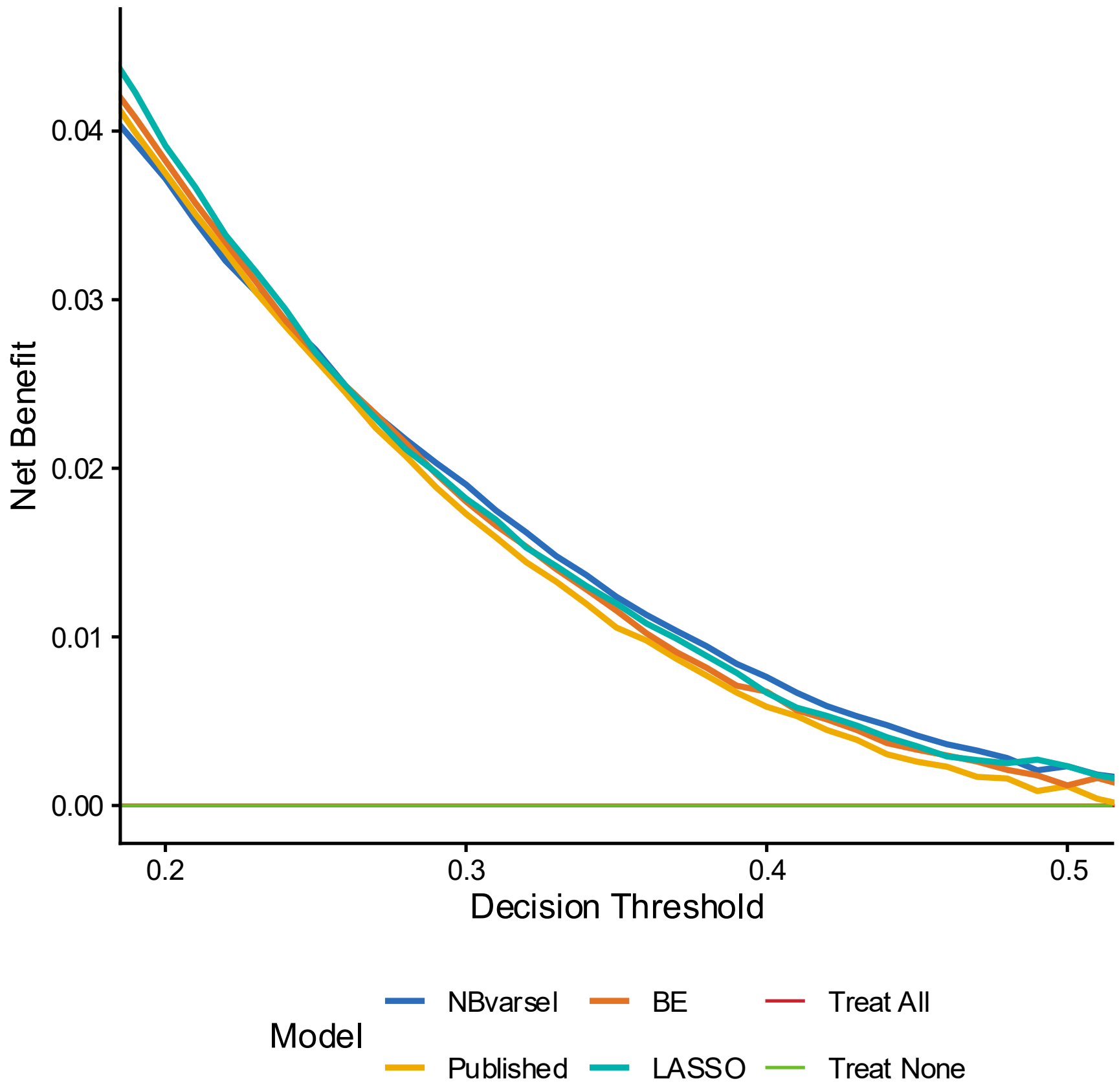

**Figure 3. Optimism corrected decision curve analysis across reasonable clinically relevant thresholds based on 1000 bootstraps (Median results).**

**Table 4. Summary of results across the case studies. Comparator is the full model for ovarian cancer and hospitalization case studies and the published model for the non-typhoidal salmonella case study. Positive Δ values refer to better performance for the NB variable selection.**

| Prediction task | Training data | Test data | Cost information | # potential predictors | # predictors NBvarsel | $\Delta$AUC | $\Delta$OE | $\Delta$NB | $\Delta$Brier |
|---|---|---|---|---|---|---|---|---|---|
| Risk of malignancy of ovarian mass [16] | 5,909 | 6,847 | Yes | 16 | 13 | -0.001 | 0.0189 | 0.0122 | -0.001 |
| Risk of hospitalization after emergency department visit [26] | 334,420 | 83,605 | No | 64 | 49 (BE 1)<br>48 (BE 2) | 0.001<br>0.001 | 0.0001<br>0.0001 | 0.0024<br>0.0022 | 0.001<br>0.001 |
| Model predicts NTS bloodstream infection and can support the choice of empiric antibiotics to include coverage of drug-resistant NTS [27] | 2323 | 1000 bootstraps | No | 11 | 6[a] | −0.005 | −0.0026 | 0.0012 | 0.000 |

$\Delta$OE: Difference between absolute distance from perfect calibration of comparator model and NBvarsel model.

$\Delta$NB: Difference between NBvarsel across clinically relevant thresholds and comparator.

$\Delta$Brier: Difference in Brier score between comparator model and NBvarsel model.

[a]6 predictors were selected in the apparent model but across bootstraps the mean was 7.

# 5 DISCUSSION

## 5.1 SUMMARY OF KEY FINDINGS

While classical statistical properties are widely used to build clinical prediction models, relying solely on statistical metrics for variable selection often fails to optimise clinical utility and to account for the real-world burden of measuring predictors. To address this gap, we developed and implemented a variable selection algorithm as an R package that determines the optimal subset of predictors by exhaustively exploring all predictor combinations and maximizing clinical utility, measured via harm-adjusted or traditional NB. Through our motivating example and three case studies, we showed that this decision analysis-based approach effectively reduces model complexity, maintaining comparable discrimination, calibration and improving clinical utility. With the motivating example, we showed a hypothetical situation where the algorithm will effectively remove a very costly imaging test, leading to lower discrimination but being more useful in practice because it would not require the patient to undergo a burdensome imaging test. In the ovarian cancer case study, adjusting for measurement harm removed the CA125 biomarker, a test requiring a blood draw, while maintaining comparable performance to traditional backward elimination and LASSO methods. Furthermore, we showed that by using a backward elimination approach, this methodology can be successfully scaled to datasets with many predictors like MIMIC IV-ED. The variable selection procedure showed increased utility when test harms can be specified (ovarian cancer case study) and when the range of clinically relevant thresholds is clearly defined (non-typhoidal Salmonella case study), but the improvement with respect to traditional methods was limited for the hospitalization case study (**Table 4**).

## 5.2 INTERPRETATION OF RESULTS

The results highlight that the "best" model in purely statistical terms is not necessarily the most useful model in clinical practice. Purely statistical metrics do not have the ability to balance correctly identified cases and harms of misclassifying a patient as done through NB. In addition, our approach can account for test harm as a crucial layer for variable selection. For instance, in the motivating example, ignoring test harms led to the inclusion of predictors that offered negligible improvements in NB at a disproportionate measurement cost. By translating the clinical burden into test harm, the minimum number of tests a doctor is willing to perform per true positive to justify the cost, our algorithm penalizes "expensive" predictors (**Box 1**). Even when eliciting test harms is not possible, models developed using this algorithm are naturally more parsimonious and tailored for pragmatic clinical implementation, as seen in the DENTS trial data where the NB-based model achieved a better median NB across the clinically-relevant threshold range while requiring fewer predictors.

## 5.3 CONTEXTUALIZATION WITHIN EXISTING LITERATURE

There is growing recognition in the field of clinical prediction modelling that traditional performance measures are insufficient to guarantee a model's utility [1,7]. However, embedding utility directly into the variable selection phase has historically been difficult. While Baker et al. discussed the use of NB to evaluate new markers, it was not operationalized as an automated variable selection algorithm [29]. Preliminary efforts to add utility to variable selection have struggled with the generation of utility values and the complexity of the methods [30]. Fouskakis and colleagues proposed using stochastic optimization to select the model with the best cost-adjusted Bayesian Information Criterion (BIC). The cost-adjusted BIC extends standard model selection by penalizing the total measurement cost of predictors rather than solely the number of variables included. However, implementing this method requires the specification of a cost-penalized prior, a task that proves difficult for clinicians who must quantify the trade-off between statistical gain and data collection burden. Zhao and Yu developed a cost-sensitive variable selection method based on the $l_{2,1}$-norm [31]. Their work is mainly methodological, without linking it to underlying utility of predictors. The *glmnet* package in R allows using weights that can represent misclassification costs [32]. The misclassification costs are used to estimate the model coefficients, therefore providing miscalibrated models, similar to class imbalance corrections [33]. The package also allows determining penalty factors (i.e. predictor costs). The predictor's cost in the package can only be determined in relative terms without clinical translation, and they do not provide a combined cost approach which is common in medical settings. Our approach bridges this methodological gap by using "simple" metrics such as NB and the test harm that have a clear clinical meaning, providing a much more intuitive scale for clinicians to quantify the harm-to-benefit ratio.

## 5.4 LIMITATIONS

Despite its advantages, this methodology has several limitations. First, determining the test harm of a predictor is subjective, highly context-dependent, and challenging to quantify precisely prior to model development (**Box 1**). When setting test harms is not possible, the algorithm can be safely used without costs and provide a variable importance metric based on NB, although the expected improvement with respect to traditional approaches is reduced. Second, this is a data-driven method and therefore it is prone to overfit if sample size is limited. We recommend using the algorithm at least with the minimum sample size recommended by Riley's method [13]. Conducting first an a-priori variable selection is also recommended to reduce the number of potential predictors [5,6]. Third, the exhaustive evaluation of all predictor combinations is computationally intensive and becomes infeasible with many candidate predictors. This is a process that should be done only once for model development, and we provide an option to parallelize the process to reduce considerably the running time as well as a backward elimination alternative. Lastly, we only developed the variable selection methodology for logistic regression, but we aim to expand to other modelling algorithms as part of the *NBvarsel* package.

## 5.5 IMPLICATIONS

Prediction models that require burdensome, invasive, or expensive data collection for only marginal statistical gains are rarely adopted in routine care [34]. By integrating clinical utility directly into the model training pipeline, developers can ensure that their final models are both statistically rigorous and practically viable. The *NBvarsel* R package available on CRAN (https://lasaibarrenada.github.io/NB_varsel/) and its associated visualizations provide researchers with accessible tools to make informed, utility-driven decisions about variable selection.

## 5.6 FUTURE RESEARCH

Future research should prioritize extensive, controlled evaluations of this methodology. Specifically, simulation studies are needed to determine the sample sizes required to prevent overfitting, compare the relative performance of exhaustive versus BE approaches, and assess the impact of threshold mismatches between training and testing phases. We also plan to expand the methodology's analytical capabilities. For the current binary logistic regression framework, future updates will introduce full control over variable interactions, spline terms and the possibility of adding conditional test harms. Furthermore, while we currently use the average NB across a clinically relevant threshold range, we aim to implement a weighted average based on the distance to the optimal threshold. In this line, allowing the algorithm to account for multiple decisions across the different decision thresholds and optimize the continuous net benefit could be an addition for cases where a model is intended to support multiple decisions with multiple thresholds [35]. Another related approach to using clinical utility directly into parameter estimation would involve using harm-adjusted NB as the optimization criterion for a loss function in deep learning architectures [36]. Although the algorithm's fundamental rationale applies to any prediction model, the current software is limited to binary logistic regression; future iterations will support survival and time-to-event outcomes, alongside other modelling algorithms. Finally, this package is open-source and intended for continuous development, and we actively welcome contributions from the research community.

# 6 CONCLUSION

Clinical utility should not be an afterthought reserved only for external validation. By integrating a harm-adjusted NB metric directly into the variable selection process, researchers can produce prediction models that maximize patient benefit while minimizing unnecessary clinical burden. We present an easy-to-use algorithm based on the utility a model would have in the clinic to select the best subset of predictors, accounting for cost of measuring those, which leads to simpler models maintaining or improving the utility.

**Author contributions**

Conceptualization: LB, AV, LW, BVC, ES. Methodology: LB, AV, LW, BVC, ES. Investigation: LB, AV, LW, BVC, ES. Visualization: LB. Funding acquisition: BVC, DT, LB. Project administration: LB. Supervision: AV, LW, BVC, ES. Writing – original draft: LB, AV. Writing – review & editing: LB, AV, LW, BVC, ES, DT, BT.

**Funding**

Financial support for this study was provided by the Research Foundation – Flanders (FWO) under grant V457024N and 12A8F27N for LB and G097322N for BVC and DT. This work was supported in part by the National Institutes of Health/National Cancer Institute (NIH/NCI) with a Cancer Center Support Grant to Memorial Sloan Kettering Cancer Center [P30-CA008748] for AV. The funding agreements ensured the authors' independence in designing the study, interpreting the data, writing, and publishing the report. LW is supported by the VIDI talent programme of the Dutch Research Council (NWO) (file number 09150172310023)

**Conflicts of interest**

Nothing to declare

**Data availability**

The IOTA data for the ovarian cancer case study is not publicly available nor the non-typhoidal Salmonella case study from the DENTS trial. The MIMIC IV (version 3.1) and MIMIC ED (version 2.2) are publicly available but due to patient privacy and the PhysioNet Data Use Agreement, the raw data cannot be shared directly by the authors. Access to the database is freely available to researchers who complete the required human-subjects research training (e.g., CITI training) and formally apply for credentialed access via PhysioNet. The code used for the analysis of the case studies is provided in the package documentation for transparency and replicability.

# Cost-based variable selection for prediction models

Lasai Barreñada[1,2], Andrew Vickers[3], Laure Wynants[1,2,4], Bieke Tack[5], Dirk Timmerman[1,6], Ewout Steyerberg[7], Ben Van Calster[1,2,7]

# SUPPLEMENTARY MATERIAL

[1] Department of Development and Regeneration, KU Leuven, Belgium

[2] Leuven Unit for Health Technology Assessment Research (LUHTAR), KU Leuven, Belgium

[3] Department of Epidemiology and Biostatistics, Memorial Sloan Kettering Cancer Center, New York, NY, USA

[4] Department of Epidemiology, CAPHRI Care and Public Health Research Institute, Maastricht University, Maastricht, Netherlands

[5] Department General Internal Medicine, University Hospitals Leuven, Leuven, Belgium

[6] Department Obstetrics & Gynecology, University Hospitals Leuven, Leuven, Belgium.

[7]Julius Centre for Health Sciences and Primary Care, University Medical Centre Utrecht, Utrecht, Netherlands

**Corresponding author**

Ben Van Calster

KU Leuven, Department of Development and Regeneration

Herestraat 49 box 805 T +32 16 377 788

3000 Leuven

Belgium

ben.vancalster@kuleuven.be

# SUPPLEMENTARY MATERIALS

We recommend exploring the supplementary materials in the **Online Supplementary report including also the online only figures and tables** (https://github.com/LasaiBarrenada/NB_varsel/blob/v0.1.2/NBvarsel_supplementary_report.html) with higher resolution and the whole process includes the necessary code to replicate the results.

## 1.1 INDEX OF SUPPLEMENTARY AND ONLINE ONLY FIGURES

| Identifier | Description | Study / Context |
|---|---|---|
| **Figure S1** | Permutation variable importance across top 10% models | Ovarian Cancer (IOTA) |
| **Figure S2** | Clustered calibration plot (2MA-C method) | Ovarian Cancer (IOTA) |
| **Figure S3** | All-subset model performance plot (unadjusted Net Benefit) | Ovarian Cancer (IOTA) |
| **Figure S4** | All-subset model performance plot (harm-adjusted Net Benefit) | Ovarian Cancer (IOTA) |
| **Figure S5 (left)** | Net benefit variable importance (BE NBvarsel 1) | MIMIC-IV ED |
| **Figure S5 (right)** | Net benefit variable importance (BE NBvarsel 2) | MIMIC-IV ED |
| **Figure S6** | Test-set calibration curves across comparison models | MIMIC-IV ED |
| **Figure S7** | Net benefit permutation variable importance bar chart | Salmonella (DENTS) |
| **Figure S8** | All-subset model performance plot | Salmonella (DENTS) |
| **Figure S9** | Predicted risk distributions by outcome (NBvarsel vs Published) | Salmonella (DENTS) |
| **Figure S10** | Apparent calibration curves | Salmonella (DENTS) |
| **Figure S11** | Apparent decision curve analysis | Salmonella (DENTS) |
| **Figure S12** | Predictor retention frequencies across 1,000 bootstrap iterations | Salmonella (DENTS) |
| **Online Only Figure 1** | Net benefit-based variable importance for the simple illustration | Motivating Example |
| **Online Only Figure 2** | Predicted risk distributions by malignancy status | Ovarian Cancer (IOTA) |
| **Online Only Figure 3** | Scatter plots comparing BE NBvarsel vs full model predicted risks | MIMIC-IV ED |
| **Online Only Figure 4** | Histograms of predicted probabilities by hospitalization status | MIMIC-IV ED |
| **Online Only Figure 5** | Decision curve analysis comparing net benefit across comparison models | MIMIC-IV ED |
| **Online Only Figure 6** | Histograms of estimated risk distributions by NTS infection status | Salmonella (DENTS) |
| **Online Only Figure 7** | Optimism-corrected decision curve analysis across 0%–100% thresholds | Salmonella (DENTS) |

## 1.2 INDEX OF SUPPLEMENTARY & ONLINE ONLY TABLES

| Identifier | Description | Study / Context |
|---|---|---|
| **Table S1** | Predictor and outcome distributions by training/test set | Ovarian Cancer (IOTA) |
| **Table S2** | Predictor and outcome distributions by training/test set | MIMIC-IV ED |
| **Table S3** | Descriptive baseline statistics | Salmonella (DENTS) |
| **Table S4** | Bootstrap discrimination, calibration, and Brier metrics | Salmonella (DENTS) |
| **Online Only Table 1** | Minimum sample size calculation (Riley criteria) | Ovarian Cancer (IOTA) |
| **Online Only Table 2** | Best model statistics from exhaustive search | Ovarian Cancer (IOTA) |
| **Online Only Table 3** | Full performance table of all evaluated subset models | Ovarian Cancer (IOTA) |
| **Online Only Table 4** | Decision-threshold adjusted net benefit table | Ovarian Cancer (IOTA) |
| **Online Only Table 5** | Minimum sample size calculation (Riley criteria) | MIMIC-IV ED |
| **Online Only Table 6** | Detailed discrimination and calibration performance metrics | MIMIC-IV ED |
| **Online Only Table 7** | Minimum sample size calculation (Riley criteria) | Salmonella (DENTS) |
| **Online Only Table 8** | Predictor retention percentages across bootstrap iterations | Salmonella (DENTS) |
| **Online Only Table 9** | Optimism-corrected median net benefit across 20%–50% thresholds | Salmonella (DENTS) |

## S1. VARIABLE IMPORTANCE

Variable importance is typically assessed using the statistical properties of the model being analysed. For instance, in a logistic regression model, coefficients can be used to compare the relative contribution of each predictor, while in random forest the Gini importance is often used to rank predictors by their contribution to the reduction of impurity. However, these techniques assess variable importance based on predicted probabilities rather than the clinical utility of the model. In his seminal paper introducing random forests, Breiman introduced permutation variable importance [37]. The algorithm proceeds as follows:

1. A model is trained on training data.
2. The performance of the model is evaluated on test or unobserved data.
3. The performance of the model on test data is recalculated by randomly shuffling (permuting) values of a predictor $P$.
4. The variable importance of P is the difference between the performance on the true dataset and on the permuted dataset.

We apply this approach to calculate the average contribution of each predictor to the NB across the different predictor combinations evaluated. The permuted performance is calculated within the same $k$ folds of the cross-validation to ensure that the permuted and original performances are comparable. Then this importance can be summarised across all or a subset of combinations.

## S2. SIMULATION STUDY FOR THE MOTIVATING EXAMPLE

As a motivating example, we simulated a dataset ($N = 10{,}000$) to evaluate a biopsy risk prediction model incorporating three continuous predictors (mean = 0, SD = 1), with the binary outcome generated via Bernoulli trials based on the true data-generating model, resulting in a 22% prevalence. The true data-generating model for the outcome is defined as $logit(Y) = -1.5 + 0.8X_{img} + 0.5X_{blood} + 0.4X_{clin}$. In this scenario, $X_{img}$ represents a strongly predictive but highly burdensome imaging test (test harm = 0.05), $X_{blood}$ is a moderate blood biomarker (test harm = 0.01), and $X_{clin}$ is a weakly predictive, cost-free clinical variable. These simulation factors were purposefully selected to demonstrate a scenario where a moderately discriminative, low-cost model is preferable to a highly discriminative but costly one.

## S3. Non-typhoidal Salmonella case study apparent performance.

According to NBvarsel, the model removed age, nutritional status, haemoglobin levels, respiratory distress and antibiotics use and the most important predictors were malaria status, duration of fever and respiratory rate (**Figures S7-S8**). Individual risks varied considerably among the published model and the NBvarsel model (**Figure S9**). AUC was slightly higher for the published model and BE ranging for all methods between 0.79 to 0.80. Both the published model and LASSO showed good apparent calibration across the risk range. The calibration was closer to the diagonal for NBvarsel model for estimated risk below 45%. NBvarsel and BE tended to estimate too low risks in higher estimated probabilities whereas the published model showed good calibration (**Figure S10**). The NBvarsel model had higher NB from decision thresholds of 25% till 55% (**Figure S11**).

# S4. SUPPLEMENTARY TABLES AND FIGURES

**Table S1. Predictor and outcome distribution of the ovarian cancer case study by training/test.**

| | **Training** *IOTA 1-3* | | **Test** *IOTA 5* | |
|---|---|---|---|---|
| **Characteristic** | **Benign** N = 3,980 (67%) | **Malignant** N = 1,929 (33%) | **Benign** N = 5,134 (75%) | **Malignant** N = 1,713 (25%) |
| Age | 42 (32, 54)[8-96] | 57 (46, 66)[14-94] | 47 (35, 61)[18-99] | 57 (46, 67)[18-92] |
| CA125 | 18 (11, 39)[1-40,140] | 181 (37, 647)[2-38,161] | 16 (9, 33)[1-36,620] | 133 (34, 538)[3-78,200] |
| Missing | 1,447 (36%) | 358 (19%) | 2,998 (58%) | 228 (13%) |
| Family history of OC | 79 (2.0%) | 85 (4.4%) | 158 (3.1%) | 76 (4.4%) |
| >10 Locules | 199 (5.0%) | 272 (14%) | 307 (6.0%) | 327 (19%) |
| Oncology Center | 2,178 (55%) | 1,571 (81%) | 2,926 (57%) | 1,449 (85%) |
| Max. diam. lesion | 63 (45, 87)[8-550] | 89 (58, 133)[8-430] | 51 (36, 73)[7-751] | 86 (56, 125)[10-510] |
| N. papillations | | | | |
| 0 | 3,424 (86%) | 1,347 (70%) | 4,623 (90%) | 1,244 (73%) |
| 1 | 333 (8.4%) | 162 (8.4%) | 310 (6.0%) | 154 (9.0%) |
| 2 | 80 (2.0%) | 68 (3.5%) | 87 (1.7%) | 56 (3.3%) |
| 3 | 66 (1.7%) | 71 (3.7%) | 47 (0.9%) | 48 (2.8%) |
| >4 | 77 (1.9%) | 281 (15%) | 67 (1.3%) | 211 (12%) |
| Acoustic shadows | 676 (17%) | 66 (3.4%) | 1,179 (23%) | 125 (7.3%) |
| Ascites | 64 (1.6%) | 656 (34%) | 54 (1.1%) | 436 (25%) |
| Irregular walls | 1,130 (28%) | 1,245 (65%) | 1,171 (23%) | 1,078 (63%) |
| Bilateral | 540 (14%) | 601 (31%) | 718 (14%) | 567 (33%) |
| Colour score | | | | |
| 1 | 1,565 (39%) | 70 (3.6%) | 2,781 (54%) | 56 (3.3%) |
| 2 | 1,493 (38%) | 411 (21%) | 1,483 (29%) | 280 (16%) |
| 3 | 788 (20%) | 849 (44%) | 754 (15%) | 783 (46%) |
| 4 | 134 (3.4%) | 599 (31%) | 116 (2.3%) | 594 (35%) |
| Pain | 835 (21%) | 257 (13%) | 792 (15%) | 196 (11%) |
| Max. diam. solid part | 0 (0, 12)[0-50] | 50 (30, 50)[0-50] | 0 (0, 0)[0-751] | 49 (26, 75)[0-250] |
| Presence of papilations | 556 (14%) | 582 (30%) | 511 (10.0%) | 469 (27%) |
| Proportion of solid tissue | 0 (0, 0.20)[0-1] | 0.45 (0.30, 0.69)[0-1] | 0 (0, 0)[0-1] | 0.65 (0.37, 1)[0-1] |

[1]Median (Q1, Q3)[Min-Max]; n (%). OC: Ovarian cancer

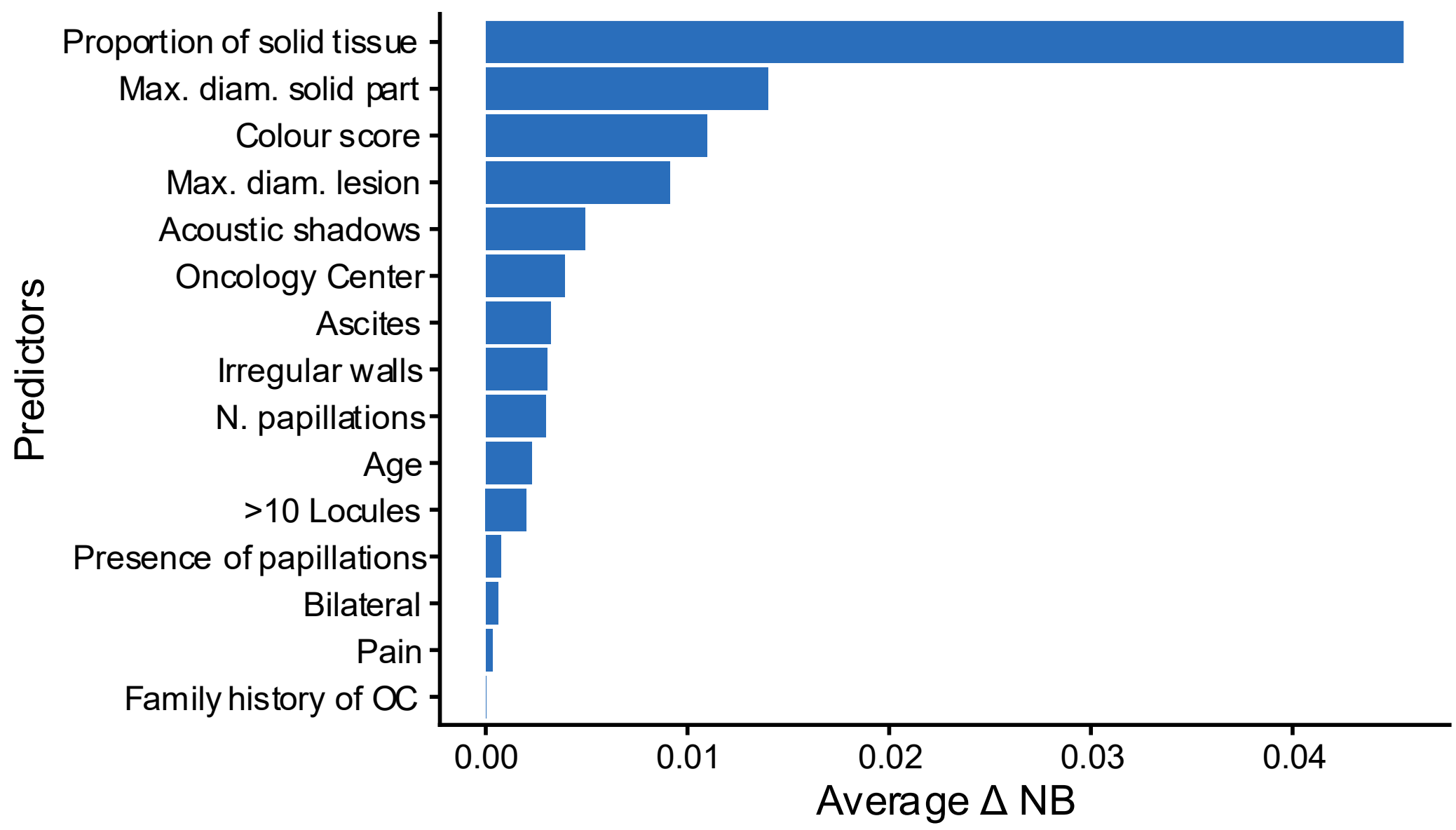

**Figure S1. Variable importance according to average improvement in NB across the 10% best performing combinations for the ovarian cancer case study.**

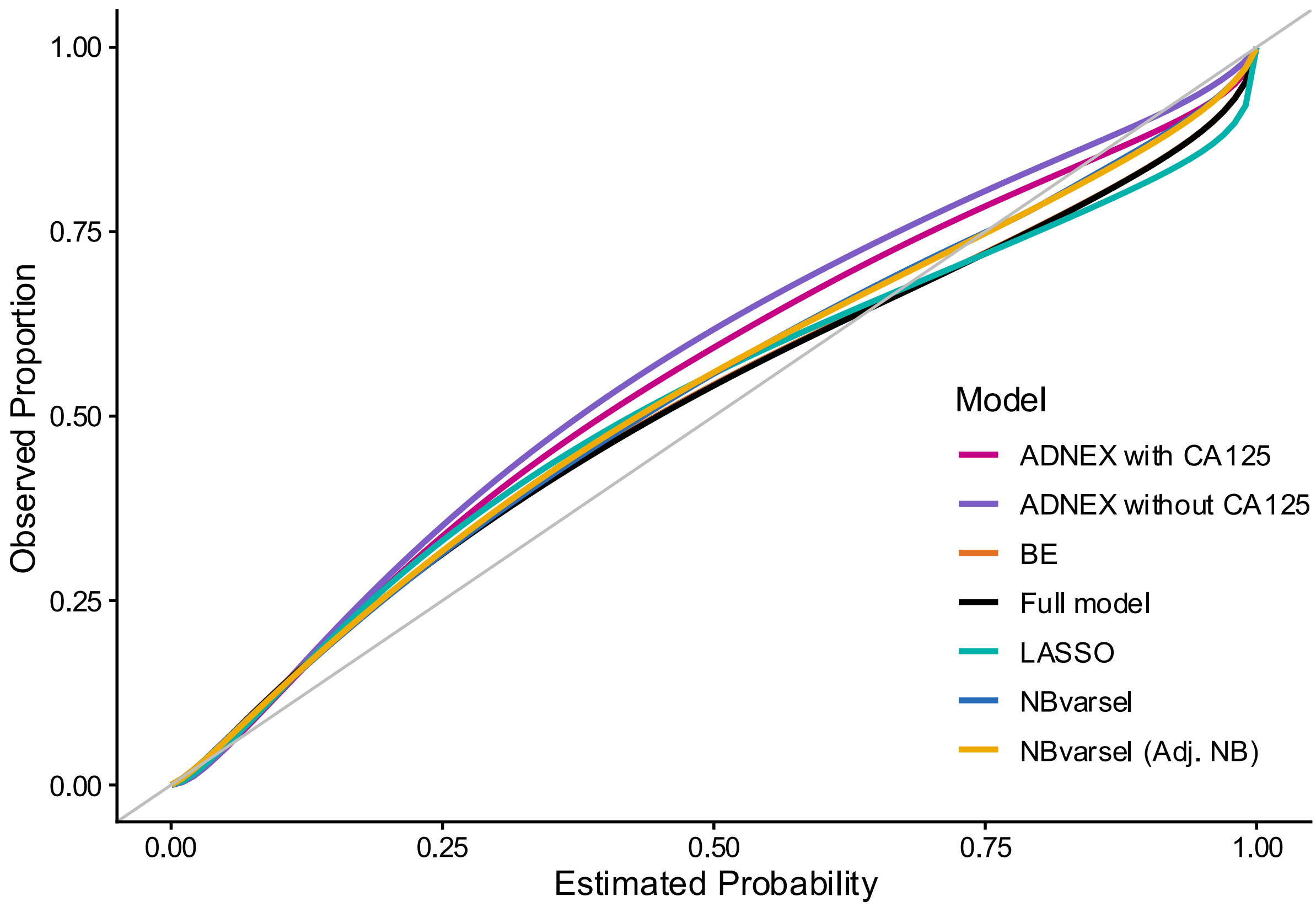

**Figure S2. Overall calibration of the clustered calibration plot for the ovarian cancer case study calculated with 2MA-C method. Histograms of predicted risks are available in the report. BE: Backward elimination; NB: Net Benefit.**

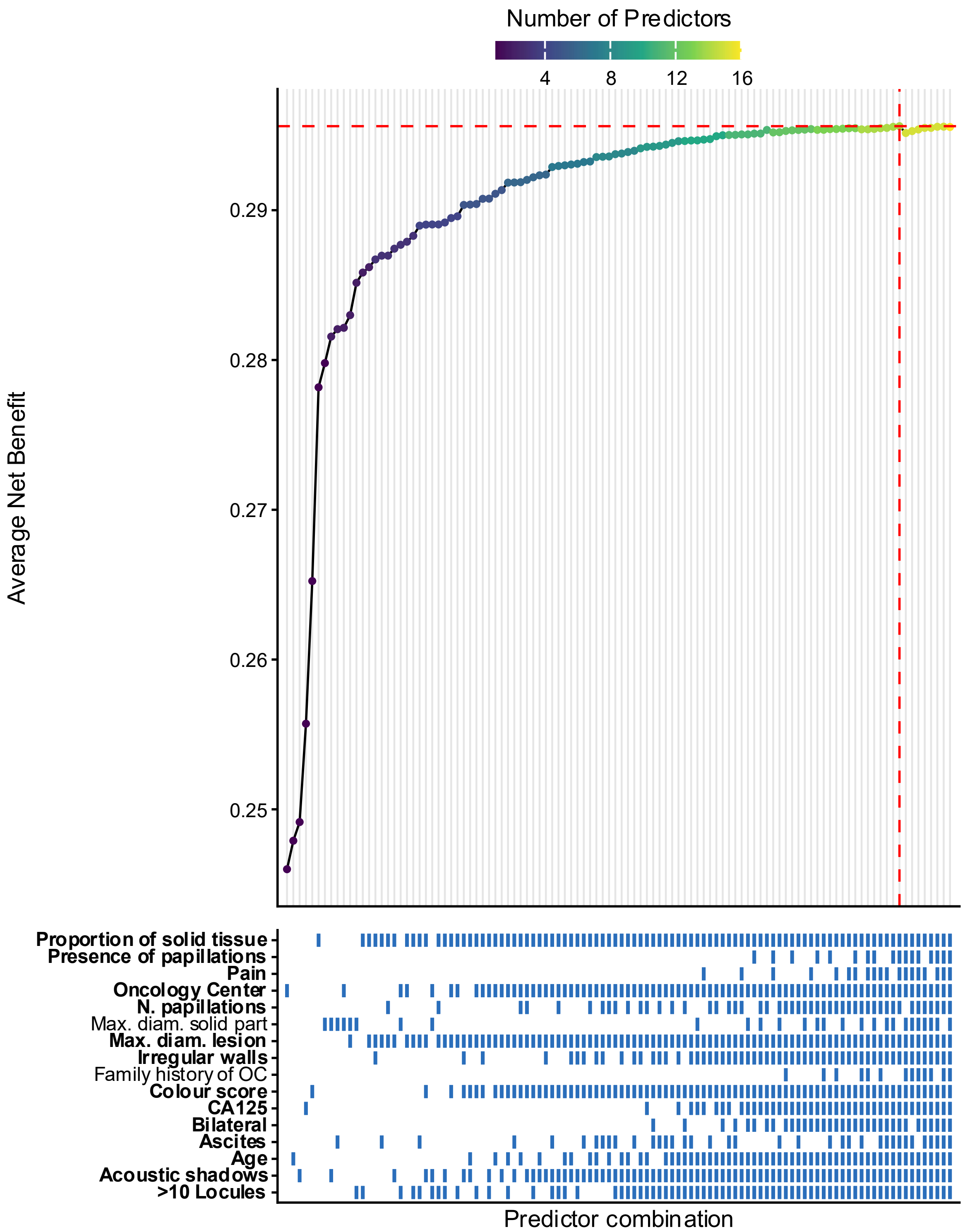

**Figure S3. All subset plot for the ovarian cancer case study ignoring harms. Terms in bold are the included predictors.**

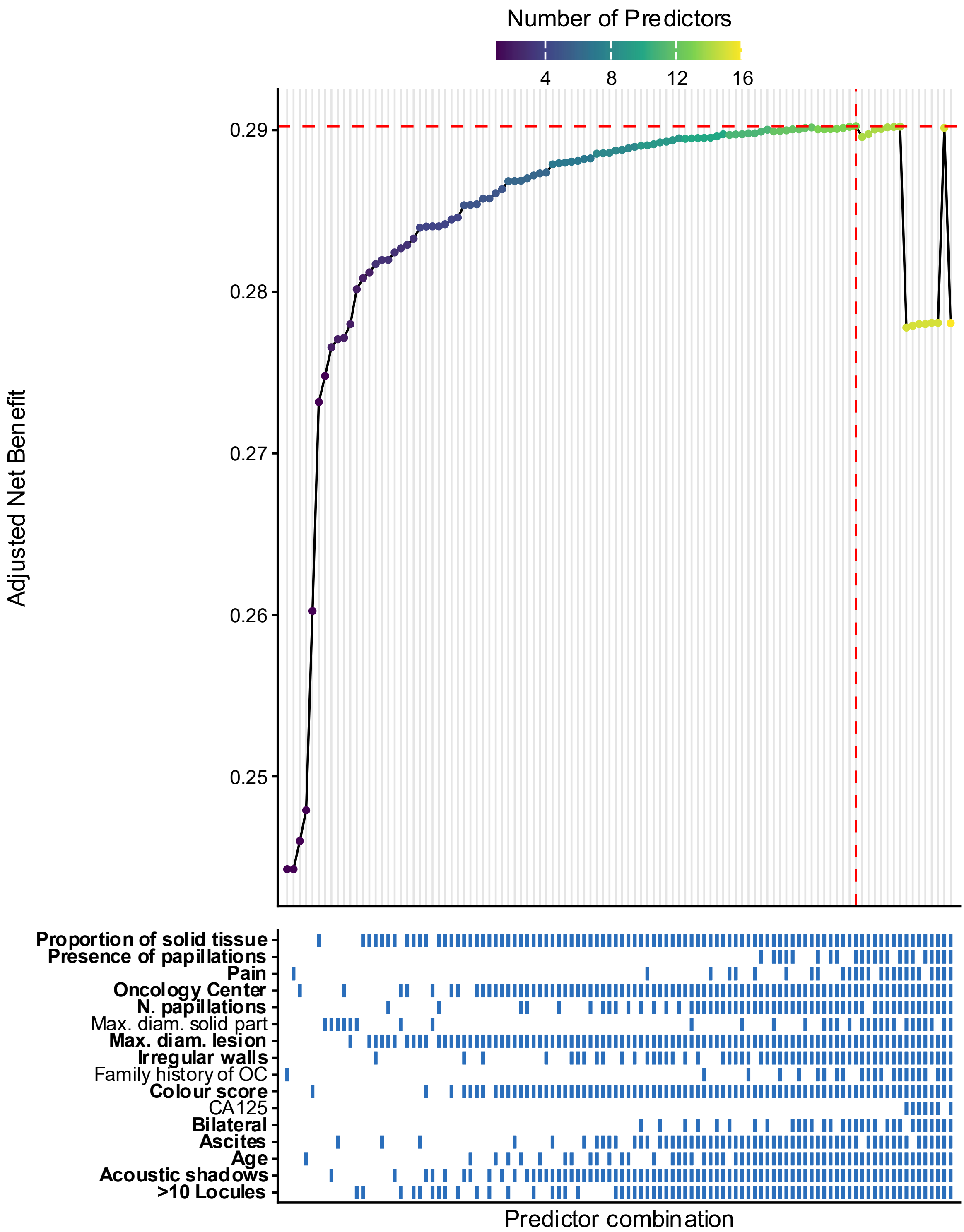

**Figure S4. All subset plot for the ovarian cancer case study accounting for harms. Terms in bold are the included predictors.**

**Table S2. Predictor distribution by outcome for the MIMIC-ED case study.**

| | Testing Set | | Training set | |
|---|---|---|---|---|
| **Predictor** | **No hosp. N = 43,957 (53%)** | **Hosp N = 39,648 (47%)** | **No hosp. N = 176,269 (53%)** | **Hosp N = 158,151 (47%)** |
| Age | 46.37 (19.40) | 60.03 (19.47) | 46.33 (19.37) | 60.09 (19.50) |
| Gender (F) | 25,284 (58%) | 20,113 (51%) | 101,441 (58%) | 80,124 (51%) |
| ED visits in the past month | 0.22 (0.81) | 0.27 (0.80) | 0.21 (0.77) | 0.27 (0.80) |
| ED visits in the past three months | 0.46 (1.64) | 0.62 (1.65) | 0.46 (1.59) | 0.62 (1.62) |
| ED visits in the past year | 1.26 (4.27) | 1.63 (4.35) | 1.25 (4.17) | 1.60 (4.24) |
| Hospitalizations in the past month | 0.09 (0.41) | 0.23 (0.60) | 0.09 (0.41) | 0.23 (0.61) |
| Hospitalizations in the past three months | 0.21 (0.82) | 0.54 (1.22) | 0.21 (0.82) | 0.54 (1.21) |
| Hospitalizations in the past year | 0.61 (2.23) | 1.42 (3.17) | 0.61 (2.21) | 1.39 (3.12) |
| ICU admissions in the past month | 0.01 (0.11) | 0.03 (0.20) | 0.01 (0.10) | 0.03 (0.20) |
| ICU admissions in the past three months | 0.02 (0.18) | 0.08 (0.35) | 0.02 (0.16) | 0.08 (0.34) |
| ICU admissions in the past year | 0.05 (0.33) | 0.19 (0.64) | 0.05 (0.30) | 0.18 (0.63) |
| Temperature (Celsius) | 36.68 (0.48) | 36.75 (0.60) | 36.68 (0.49) | 36.75 (0.58) |
| Heart rate (bpm) | 83.91 (16.33) | 86.31 (18.55) | 83.89 (16.30) | 86.31 (18.56) |
| Respiratory rate (bpm) | 17.30 (2.33) | 17.88 (2.76) | 17.30 (2.05) | 17.86 (2.85) |
| Oxygen saturation (%) | 98.81 (1.79) | 97.95 (2.80) | 98.80 (2.04) | 97.95 (2.73) |
| Systolic blood pressure (mmHg) | 135.11 (20.71) | 134.51 (23.73) | 135.18 (20.67) | 134.51 (23.68) |
| Diastolic blood pressure (mmHg) | 78.74 (13.81) | 76.01 (15.61) | 78.77 (13.74) | 76.00 (15.58) |
| Pain scale | 4.67 (3.58) | 3.59 (3.53) | 4.67 (3.59) | 3.59 (3.55) |
| Emergency Severity Index (ESI) | | | | |
| 1 | 1,003 (2%) | 3,744 (9%) | 4,061 (2%) | 15,206 (10%) |
| 2 | 8,585 (20%) | 19,365 (49%) | 34,442 (20%) | 76,998 (49%) |
| 3 | 28,666 (65%) | 16,252 (41%) | 115,259 (65%) | 64,847 (41%) |
| 4 | 5,499 (13%) | 277 (1%) | 21,649 (12%) | 1,072 (1%) |
| 5 | 204 (0%) | 10 (0%) | 858 (0%) | 28 (0%) |
| Chest Pain | 2,596 (6%) | 3,226 (8%) | 10,489 (6%) | 12,815 (8%) |
| Abdominal Pain | 4,814 (11%) | 4,718 (12%) | 19,606 (11%) | 18,996 (12%) |
| Headache | 2,191 (5%) | 856 (2%) | 9,141 (5%) | 3,546 (2%) |
| Shortness of Breath | 72 (0%) | 157 (0%) | 308 (0%) | 684 (0%) |
| Back Pain | 2,312 (5%) | 976 (2%) | 9,446 (5%) | 3,973 (3%) |
| Cough | 983 (2%) | 804 (2%) | 4,039 (2%) | 2,963 (2%) |
| Nausea/Vomiting | 1,036 (2%) | 922 (2%) | 4,280 (2%) | 3,862 (2%) |
| Fever/Chills | 857 (2%) | 2,056 (5%) | 3,553 (2%) | 8,001 (5%) |
| Syncope | 840 (2%) | 718 (2%) | 3,334 (2%) | 2,883 (2%) |
| Dizziness | 1,191 (3%) | 878 (2%) | 4,860 (3%) | 3,502 (2%) |
| Myocardial infarction | 1,195 (3%) | 3,488 (9%) | 4,997 (3%) | 13,954 (9%) |
| Congestive heart failure | 1,972 (4%) | 5,813 (15%) | 7,855 (4%) | 23,195 (15%) |
| Peripheral vascular disease | 1,083 (2%) | 3,095 (8%) | 4,385 (2%) | 12,417 (8%) |
| Stroke | 1,202 (3%) | 2,828 (7%) | 4,944 (3%) | 11,137 (7%) |
| Dementia | 369 (1%) | 1,046 (3%) | 1,571 (1%) | 4,016 (3%) |
| Chronic pulmonary disease | 4,407 (10%) | 7,598 (19%) | 17,602 (10%) | 29,854 (19%) |
| Rheumatoid disease | 568 (1%) | 1,126 (3%) | 2,290 (1%) | 4,683 (3%) |
| Peptic ulcer disease | 419 (1%) | 1,118 (3%) | 1,713 (1%) | 4,522 (3%) |
| Mild liver disease | 1,751 (4%) | 3,791 (10%) | 7,157 (4%) | 15,257 (10%) |
| Diabetes without chronic complications | 3,782 (9%) | 7,239 (18%) | 15,159 (9%) | 29,311 (19%) |
| Diabetes with complications | 1,495 (3%) | 3,810 (10%) | 6,004 (3%) | 15,442 (10%) |
| Hemiplegia | 304 (1%) | 677 (2%) | 1,205 (1%) | 2,697 (2%) |
| Moderate to severe chronic kidney disease | 2,151 (5%) | 6,084 (15%) | 8,405 (5%) | 24,193 (15%) |
| Local tumor, leukemia, and lymphoma | 1,482 (3%) | 4,018 (10%) | 5,814 (3%) | 15,682 (10%) |
| Moderate/severe liver disease | 391 (1%) | 1,299 (3%) | 1,504 (1%) | 5,312 (3%) |
| Metastatic solid tumor | 476 (1%) | 1,619 (4%) | 1,931 (1%) | 6,419 (4%) |
| AIDS | 316 (1%) | 468 (1%) | 1,196 (1%) | 1,938 (1%) |
| Cardiac arrhythmias | 3,562 (8%) | 8,250 (21%) | 14,311 (8%) | 32,344 (20%) |
| Valvular disease | 1,177 (3%) | 3,110 (8%) | 4,655 (3%) | 12,278 (8%) |
| Pulmonary circulation disorders | 1,057 (2%) | 2,889 (7%) | 4,242 (2%) | 11,041 (7%) |
| Hypertension, uncomplicated | 2,275 (5%) | 6,301 (16%) | 8,735 (5%) | 25,145 (16%) |
| Hypertension, complicated | 7,473 (17%) | 12,965 (33%) | 30,255 (17%) | 51,698 (33%) |
| Other neurological disorders | 2,062 (5%) | 4,277 (11%) | 8,431 (5%) | 16,790 (11%) |
| Hypothyroidism | 1,902 (4%) | 3,689 (9%) | 7,564 (4%) | 14,817 (9%) |
| Lymphoma | 248 (1%) | 694 (2%) | 920 (1%) | 2,639 (2%) |
| Coagulopathy | 1,560 (4%) | 4,411 (11%) | 6,346 (4%) | 17,264 (11%) |
| Obesity | 2,768 (6%) | 4,621 (12%) | 11,238 (6%) | 18,236 (12%) |
| Weight loss | 1,276 (3%) | 3,233 (8%) | 4,894 (3%) | 12,810 (8%) |
| Fluid and electrolyte disorders | 4,841 (11%) | 10,826 (27%) | 19,235 (11%) | 43,562 (28%) |
| Blood loss anemia | 307 (1%) | 820 (2%) | 1,310 (1%) | 3,341 (2%) |
| Deficiency anemia | 1,638 (4%) | 3,375 (9%) | 6,614 (4%) | 13,648 (9%) |
| Alcohol abuse | 2,311 (5%) | 4,122 (10%) | 9,384 (5%) | 16,602 (10%) |
| Drug abuse | 2,118 (5%) | 3,469 (9%) | 8,711 (5%) | 13,647 (9%) |
| Psychoses | 897 (2%) | 1,456 (4%) | 3,526 (2%) | 5,793 (4%) |
| Depression | 5,176 (12%) | 8,586 (22%) | 20,807 (12%) | 33,957 (21%) |

Results reported as mean (sd) for continuous predictors or n (%) for categorical predictors. The complete data dictionary for each predictor can be found in https://physionet.org/content/mimic-iv-ed/2.2/ and https://physionet.org/content/mimiciv/3.1/.

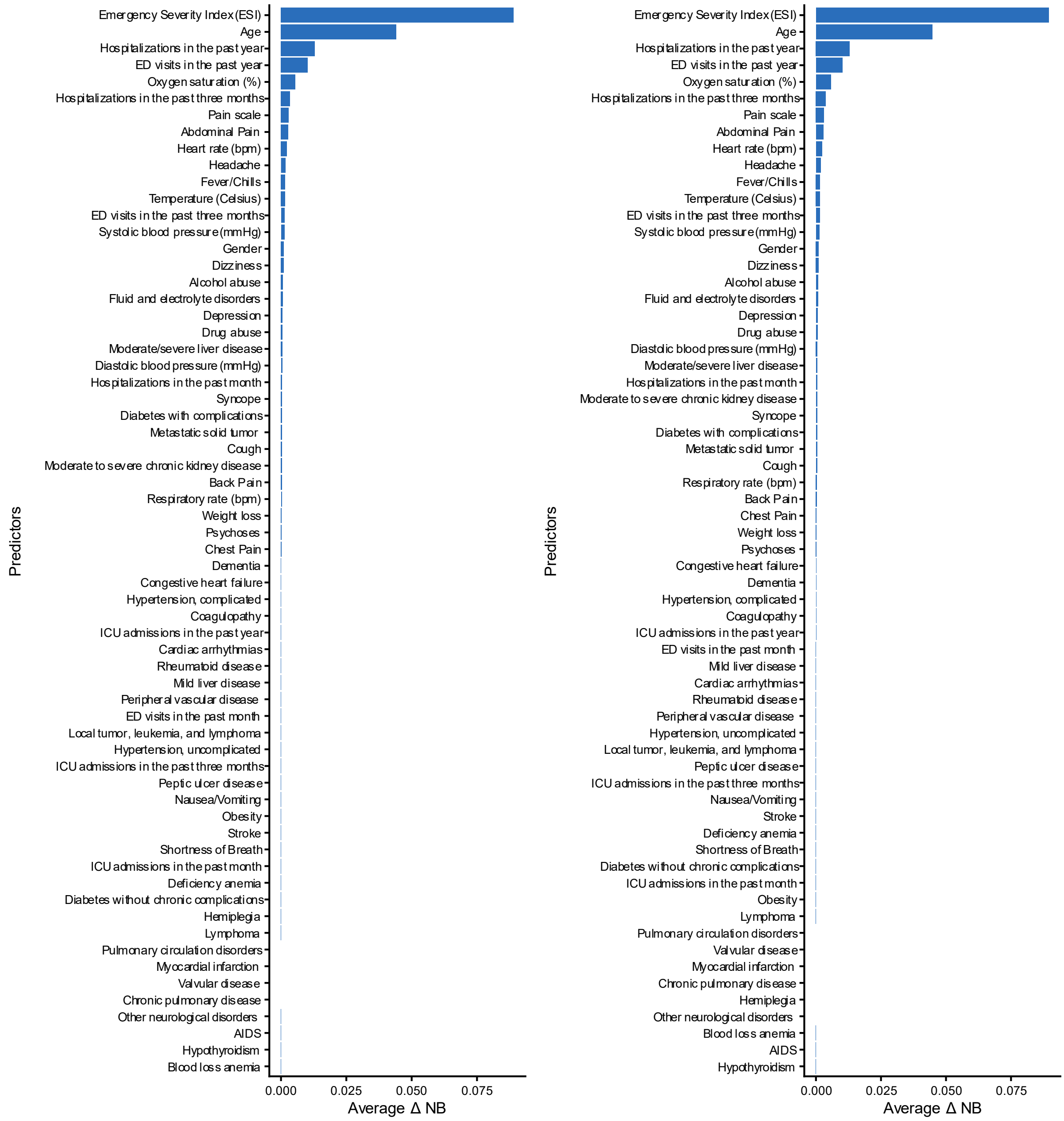

**Figure S5. Variable importance according to NB for the MIMIC-ED case study using BE NBvarsel 1 (left) and BE NBvarsel 2 (right)**

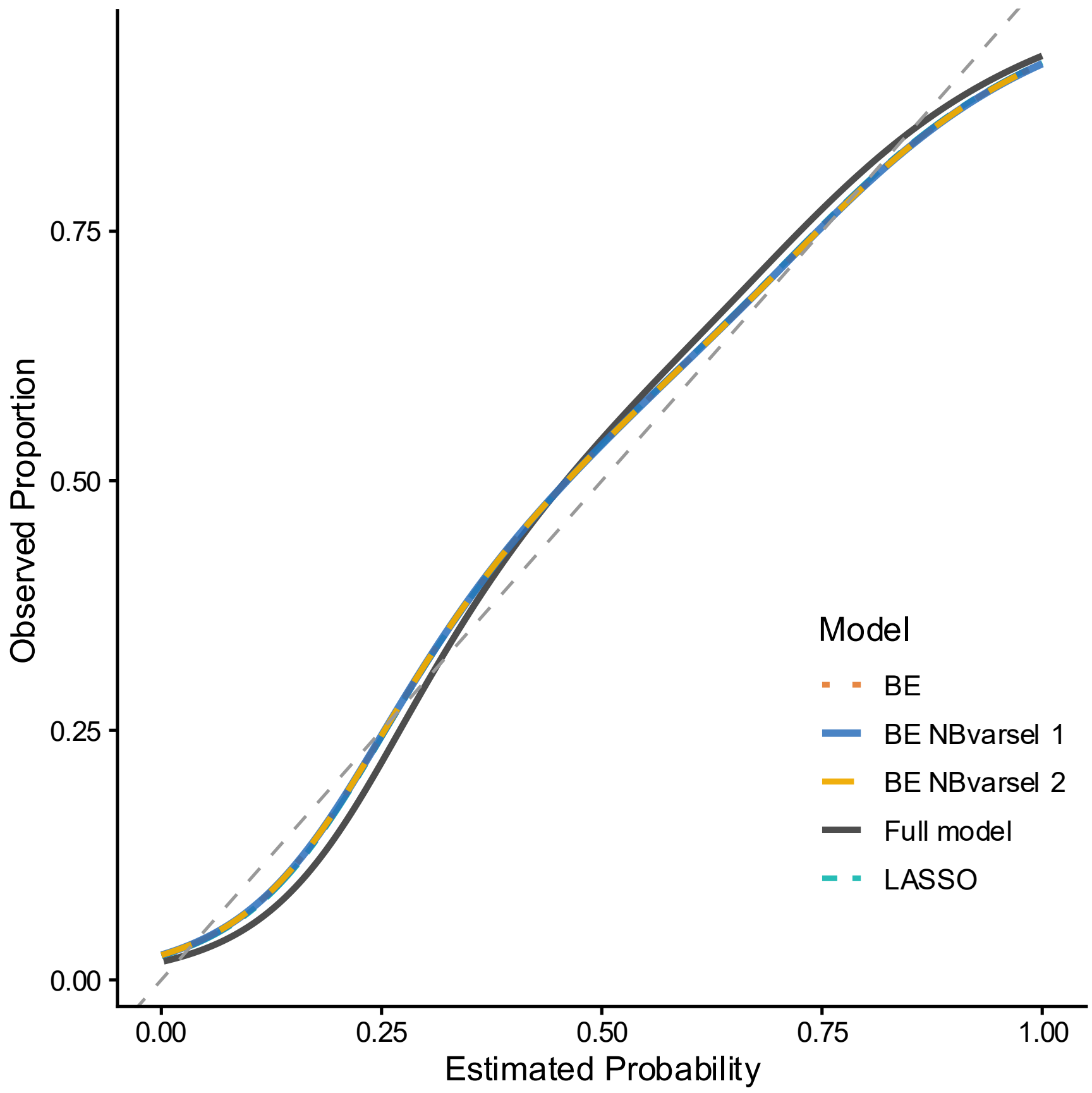

**Fig S6. Calibration plot for the MIMIC-ED case study. Calibration was virtually identical across all models.**

**Table S3. Descriptive statistics for the non-typhoidal Salmonella case study.**

| Characteristic | NTS: No<br>N = 2,029 (87%) | NTS: Yes<br>N = 294 (13%) |
|---|---|---|
| Age at admission (months) | 19 (10, 33)[1-59] | 15 (10, 23)[1-58] |
| Season of admission: Rainy | 1,416 (70%) | 245 (83%) |
| Fever duration on admission (days) | 3 (2, 4)[0-40] | 4 (3, 6)[0-19] |
| Diarrhea | 393 (19%) | 91 (31%) |
| Nutritional status | | |
| No acute malnutrition | 1,657 (82%) | 196 (67%) |
| Moderate acute malnutrition | 225 (11%) | 54 (18%) |
| Severe acute malnutrition | 147 (7.2%) | 44 (15%) |
| Respiratory rate (breaths/min) | 42 (36, 52)[6-108] | 48 (38, 59)[20-104] |
| Respiratory distress | 531 (26%) | 112 (38%) |
| Hepato- or splenomegaly | 658 (32%) | 144 (49%) |
| Malaria status | | |
| No *Pf* Malaria | 322 (16%) | 61 (21%) |
| Recent *Pf* Malaria | 318 (16%) | 148 (50%) |
| Current *Pf* Malaria | 1,389 (68%) | 85 (29%) |
| Hemoglobin (g/dl) | 7.70 (6.40, 9.30)[1.40-19.80] | 7.60 (6.30, 9.10)[2.70-17.60] |
| Antibiotics use | 721 (36%) | 128 (44%) |

[1]Median (Q1, Q3)[Min-Max]; n (%). NTS: Non-typhoidal salmonella. Pf: Plasmodium falciparum

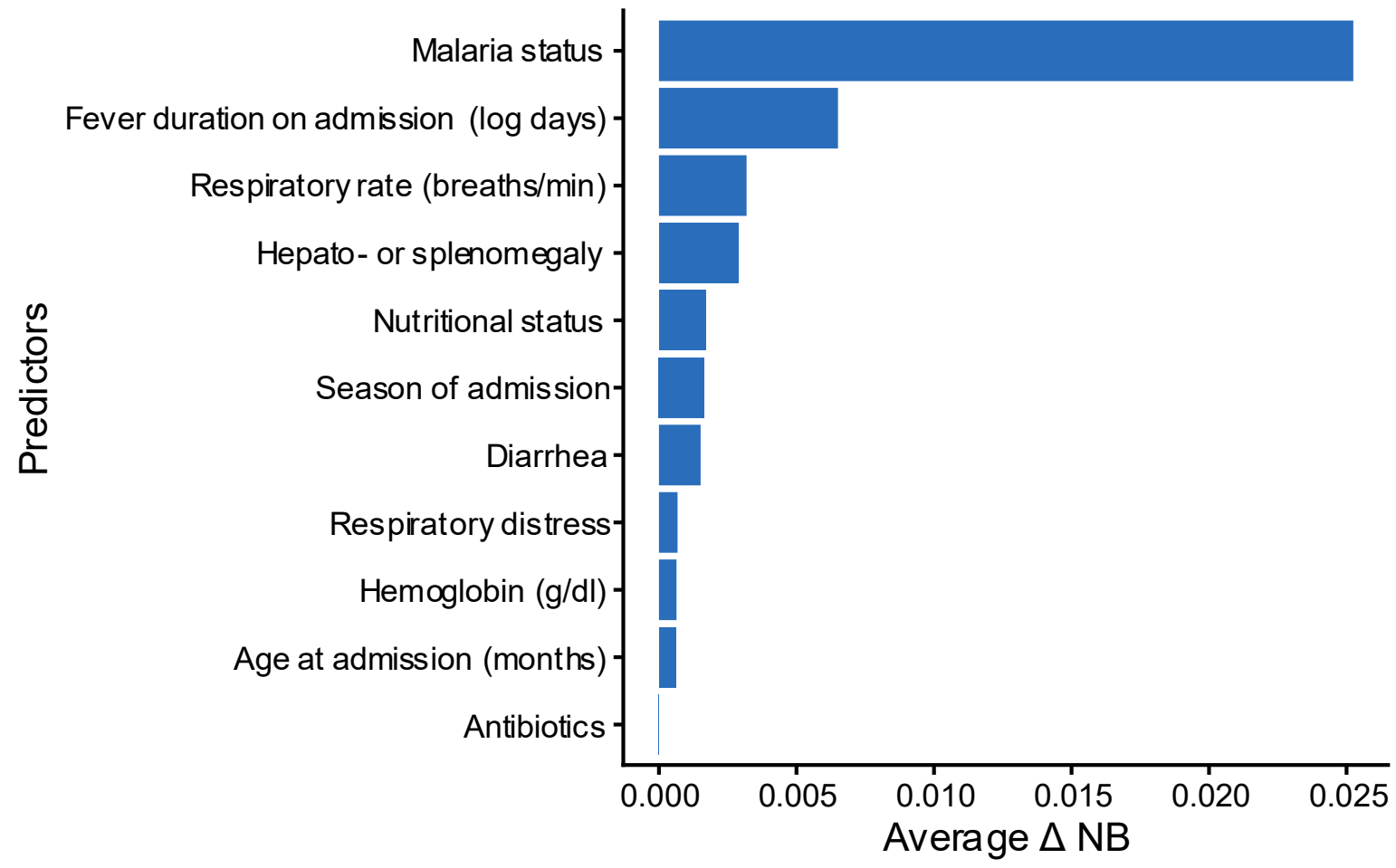

**Figure S7. Variable importance of the non-typhoidal salmonella case study.**

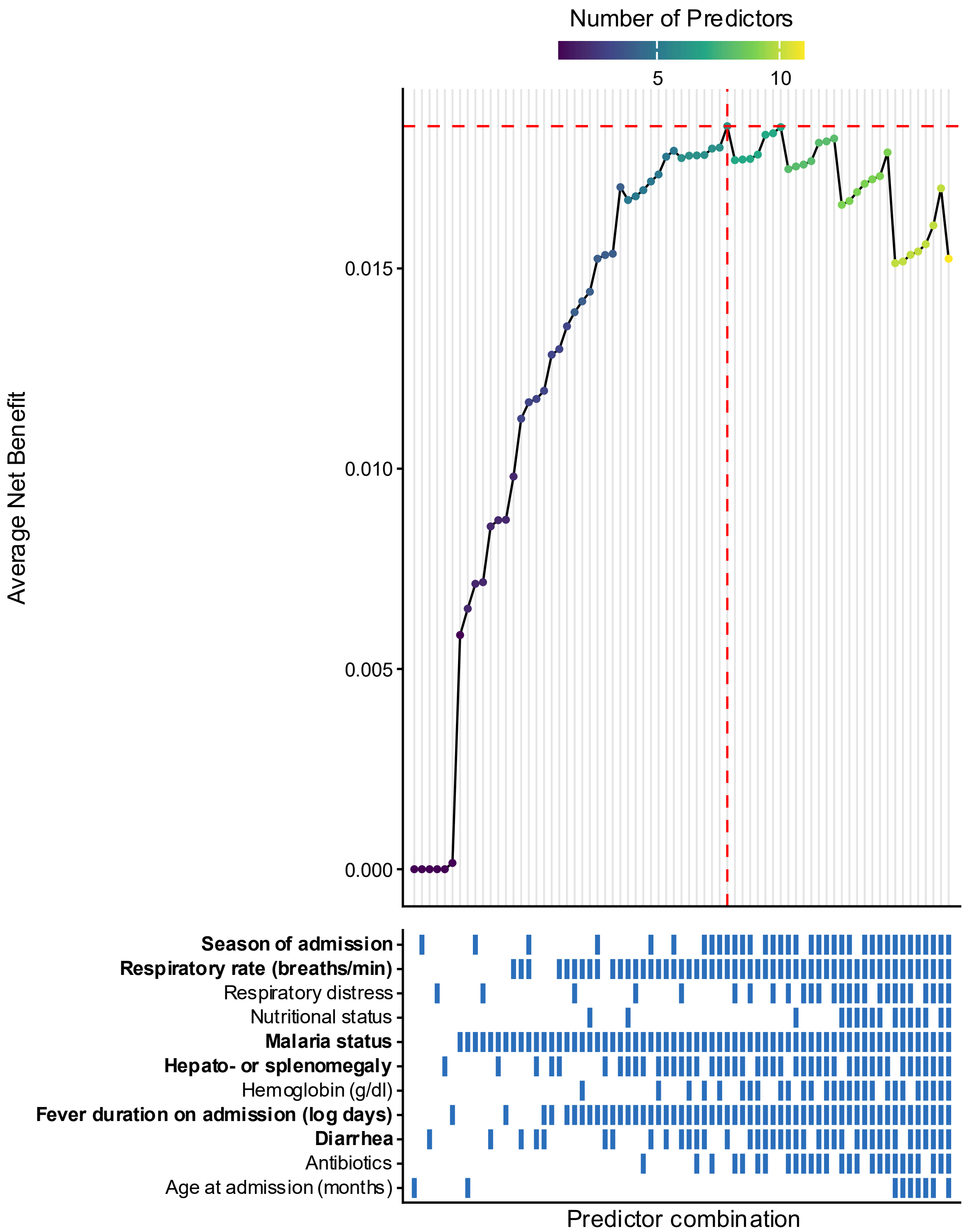

**Figure S8. All subset plot for the non-typhoidal Salmonella case study. Bold predictors are included in the final model**

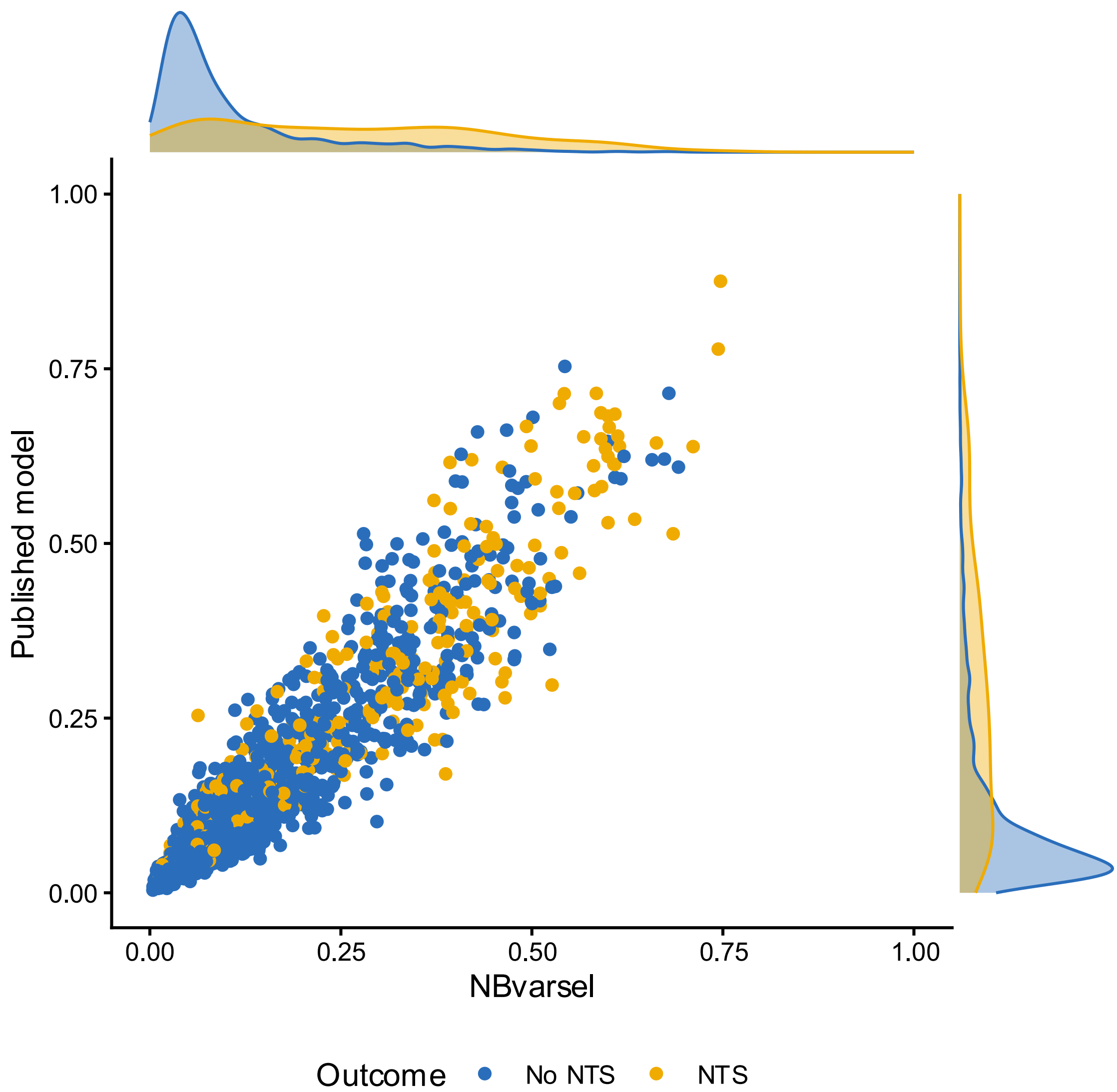

**Figure S9. Distribution of individual risks for the model based on NB variable selection and the published model stratified by outcome. Plots in the margin depict densities per outcome. NTS: non-typhoidal *S*almonella.**

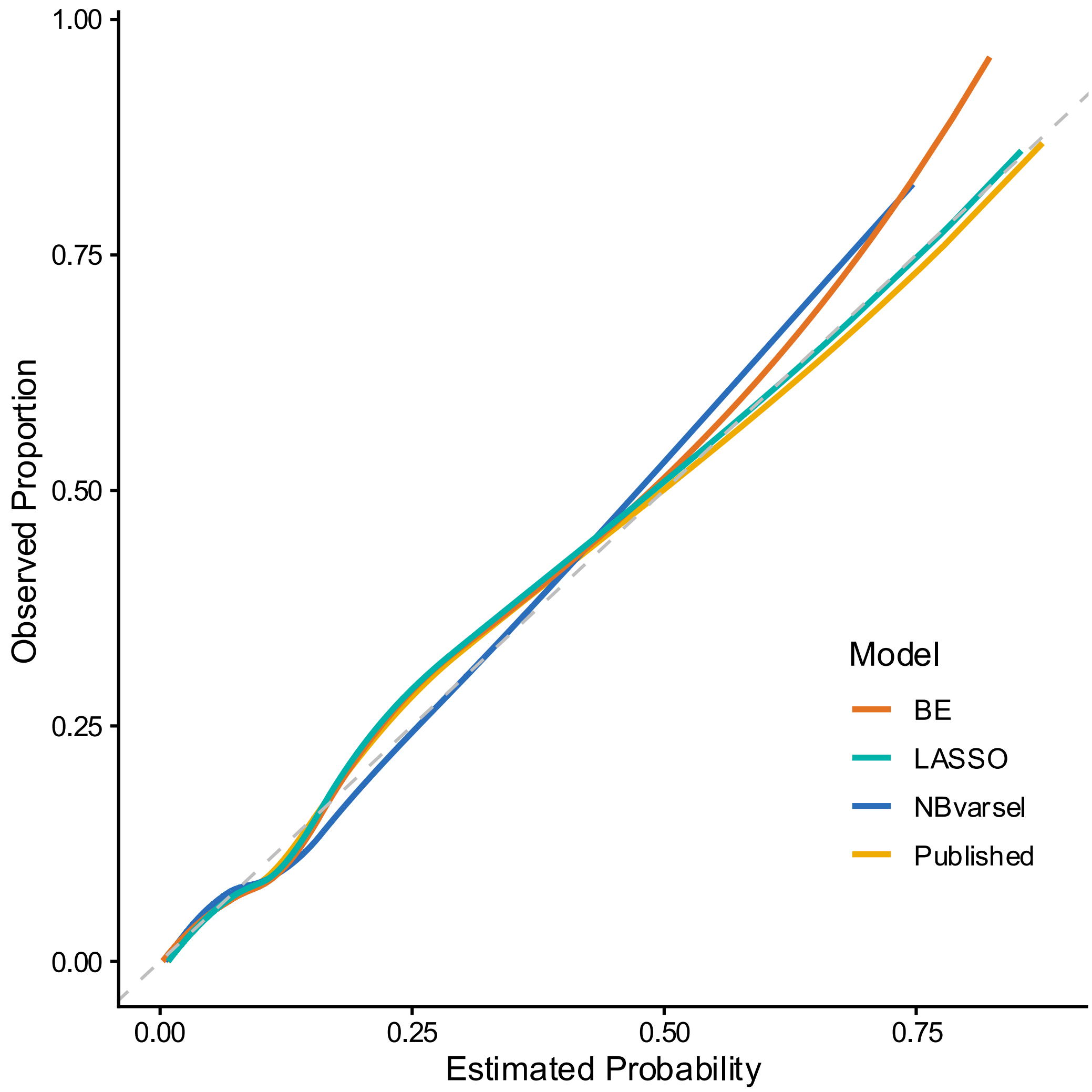

**Figure S10. Apparent calibration for the non-typhoidal Salmonella case study.**

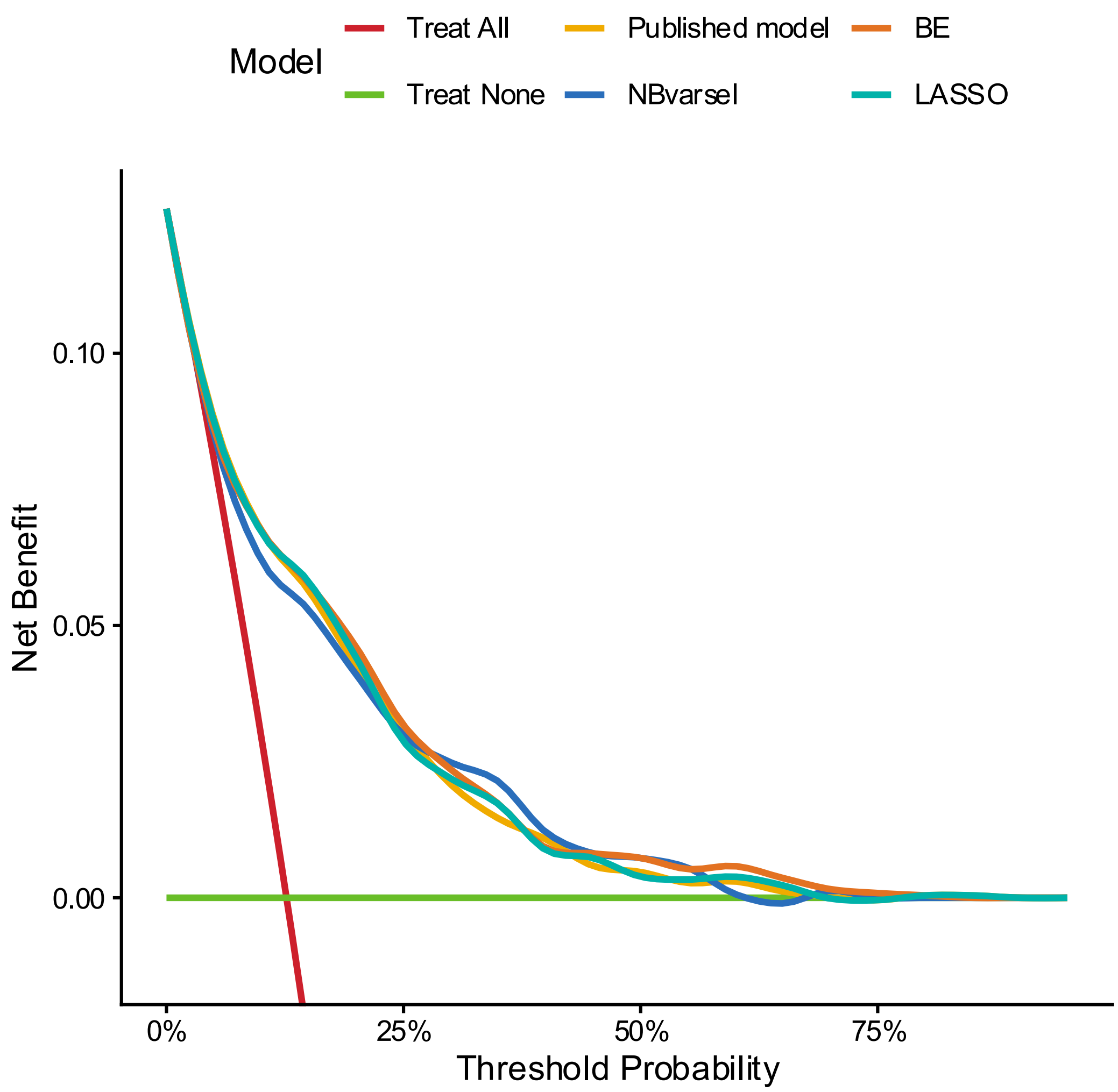

**Figure S11. Apparent decision curve analysis for the non-typhoidal Salmonella case study.**

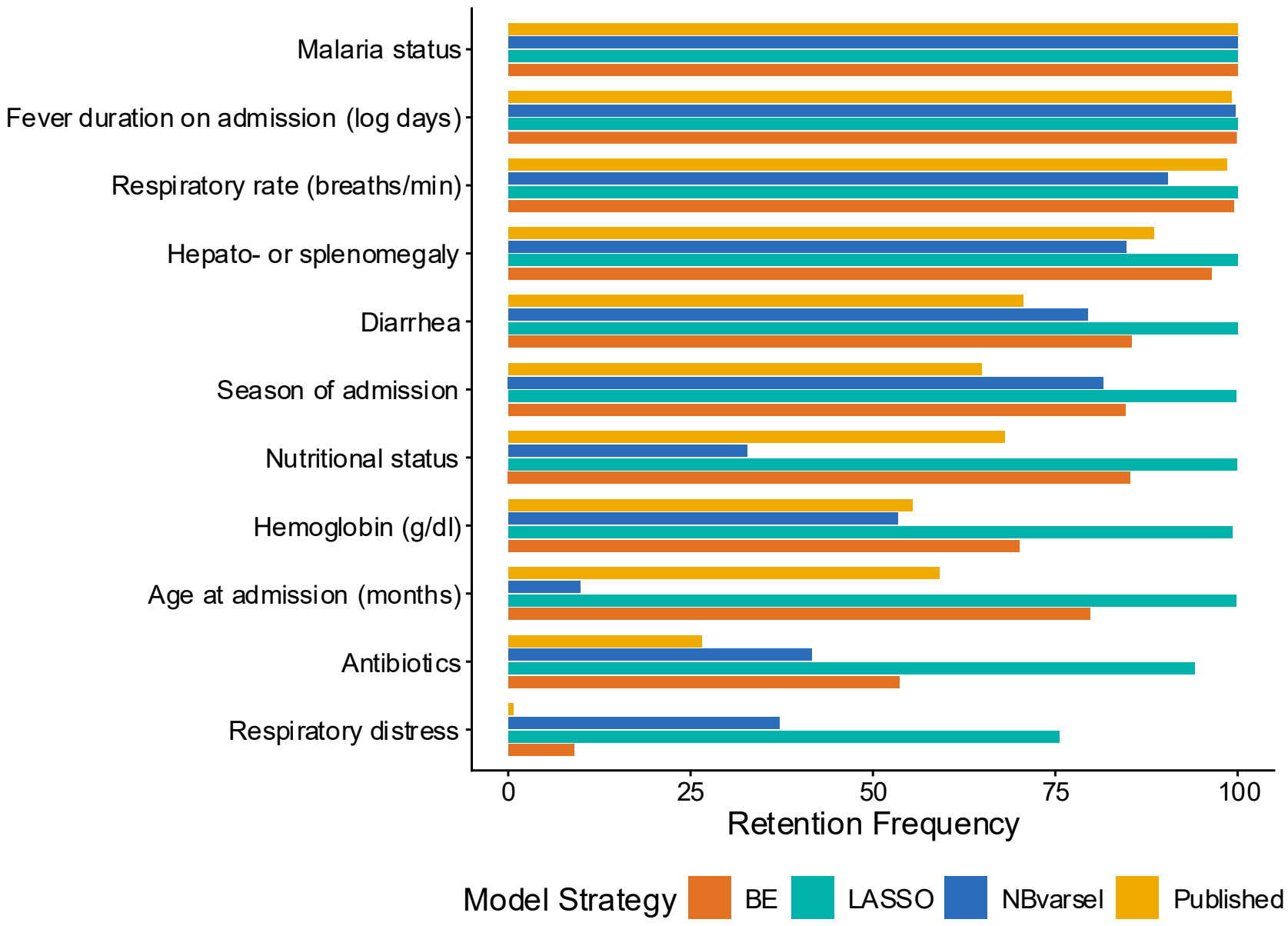

**Figure S12. Percentage of times each variable was retained across 1000 bootstrap iterations in the DENTS case study.**

**Table S4. DENTS case study bootstrap results.**

| Characteristic | NBvarsel | Published model | Backward elimination | LASSO |
|---|---|---|---|---|
| **AUC** | 0.776 (0.763-0.791) [0.696-0.841] | 0.781 (0.768-0.795) [0.714-0.835] | 0.783 (0.771-0.797) [0.713-0.840] | 0.792 (0.779-0.805) [0.715-0.844] |
| **O:E ratio** | 0.997 (0.929-1.075) [0.708-1.387] | 0.999 (0.935-1.075) [0.708-1.390] | 0.996 (0.932-1.075) [0.711-1.391] | 1.000 (0.934-1.075) [0.712-1.382] |
| **Calibration Slope** | 0.900 (0.821-0.992) [0.557-1.403] | 0.888 (0.812-0.971) [0.569-1.303] | 0.887 (0.807-0.969) [0.577-1.317] | 0.939 (0.865-1.027) [0.617-1.440] |
| **Calibration Intercept** | -0.005 (-0.100-0.099) [-0.459-0.458] | -0.001 (-0.094-0.101) [-0.457-0.457] | -0.006 (-0.098-0.103) [-0.454-0.464] | 0.000 (-0.095-0.102) [-0.450-0.455] |
| **Brier score** | 0.095 (0.091-0.099) [0.075-0.116] | 0.095 (0.092-0.099) [0.075-0.117] | 0.095 (0.091-0.099) [0.074-0.118] | 0.094 (0.090-0.098) [0.074-0.117] |

Median (Q1 - Q3) [Min - Max] across 1000 bootstrap runs.